\documentclass[aps,prl,reprint,amsmath,amssymb,longbibliography,floatfix]{revtex4-2}
\usepackage{graphicx}
\usepackage[colorlinks=true,citecolor=blue,urlcolor=blue,linkcolor=blue]{hyperref}
\makeatletter
\let\appendixsection\section
\let\appendixsubsection\subsection
\def\@hangfroms@section#1#2{#1#2}
\def\@hangfrom@section#1#2#3{\@hangfrom{#1#2}#3}

\renewcommand\section{\@startsection{section}{1}{\z@}%
  {-2.2ex plus -.5ex minus -.2ex}{1.0ex plus .2ex}%
  {\normalfont\normalsize\bfseries\raggedright}}
\renewcommand\subsection{\@startsection{subsection}{2}{\z@}%
  {-1.8ex plus -.4ex minus -.2ex}{.7ex plus .2ex}%
  {\normalfont\normalsize\bfseries\itshape\raggedright}}
\makeatother
\newcommand{\dd}{\mathrm d}
\begin{document}
\title{Non-linear dynamo waves in magnetized shear flows: a candidate mechanism for the solar cycle}
\author{Uddipan Banik}
\email{ubanik@perimeterinstitute.ca}
\affiliation{Perimeter Institute for Theoretical Physics, 31 Caroline Street N., Waterloo, Ontario, N2L 2Y5, Canada}
\date{September 2026}
\begin{abstract}
Large-scale magnetic reversals in planets, stars, and accretion disks require a non-linear theory connecting the evolving magnetic field to the flow that sustains it. To describe this coupling, we derive a family of exact non-linear solutions to incompressible magnetohydrodynamics in background rotational shear, assuming the large-scale velocity and magnetic fields to be linear in position (affine form). This reveals a cyclic feedback between meridional circulation, shear, magnetic tension, induction, and Coriolis deflection. The key solution, a steady-circulation reversal orbit or dynamo wave, periodically reverses all magnetic components, while exhibiting slower modulations. In the solar near-surface shear layer (NSSL), moderate shear permits stable 11-year reversals with poleward speeds of $8$--$12\,\mathrm{m\,s^{-1}}$ and poloidal and toroidal amplitudes of approximately $4$--$11$ G and $2.2$--$3.8$ kG near 1 Mm depth, in reasonable agreement with regional observations. These waves admit a continuous range of modulation periods encompassing the Gleissberg and Suess--de Vries bands, without preferentially selecting them. At stronger shear, a distinct 88-year reversal supports slow modulation near $203$--$207$ yr without precise shear tuning, although it requires weaker poloidal fields; an 88-year polarity reversal remains observationally unestablished however. At tachocline density, the inferred return flow and an 11-year reversal require toroidal fields of hundreds of kilogauss over $q=0.5$--$1.8$, exceeding recent large-scale-field reconstructions but partly permitted by seismic upper limits. This comparison favors the shallow NSSL for a spatially confined cycle. Our local analysis does not address the question of a global solar dynamo. Toroidal dominance prolongs the cycle because shear acting on a weak poloidal field changes only a small fraction of the stronger toroidal field per rotation. Balancing the weak magnetic torque against advection and Coriolis forces also requires slow circulation, which gradually turns the poloidal field and reverses toroidal induction over many rotations. Efficient regeneration that increases the poloidal-to-toroidal ratio would shorten this delay. The long solar cycle therefore favors a primarily shear-driven dynamo.
\end{abstract}
\maketitle
\flushbottom
\predisplaypenalty=0

\section{Introduction}
Magnetic reversals in rotating plasmas span a wide range of physical systems from planetary and stellar interiors to accretion disks. Earth's palaeomagnetic record contains irregular reversals; global rotation-convection simulations reproduce dipole-dominated fields and spontaneous reversals~\cite{ChristensenPlanetary,GlatzmaierRoberts1995}. The Sun's 11-year activity cycle (half of a 22-year magnetic cycle) shows polarity reversals over that period, along with equatorward sunspot migration (butterfly diagram) and poleward surface-field transport~\cite{Hathaway2015,Charbonneau2020}. Quasi-biennial (roughly two-year), Gleissberg, and longer activity bands show additional modulations: cycles within cycles~\cite{SolarQBO2019,Knudsen2009,Kim2020}. Accretion-disk magnetic cycles are less directly constrained, but magnetorotational-instability (MRI) simulations of angular-momentum transport through shear-driven dynamos produce long-period toroidal reversals and butterfly diagrams~\cite{Brandenburg,Hawley1996,Stone1996,Shi} unexplained by linear theory~\cite{BalbusHawley,BalbusHawleyReview}. Despite vastly different spatiotemporal scales, these systems combine rotational shear, induction, and Lorentz-force feedback. Therefore, this raises a fundamental question: is the underlying mechanism for cyclic dynamos universal?

Theory and simulations have tried to understand the large-scale dynamo in different ways. Global planetary convection models recover many observed field morphologies, although accessible diffusivities remain far from planetary values~\cite{ChristensenPlanetary}. Solar $\alpha\Omega$ theory combines poloidal-field generation by helical flows (the $\alpha$ effect) with toroidal-field generation by differential rotation, representing the turbulent electromotive force (emf) in terms of the mean field. Babcock--Leighton models of the solar dynamo rebuild the poloidal field through the emergence, tilt, transport, and decay of bipolar regions~\cite{Parker,SteenbeckKrauseRadler,Babcock1961,Leighton1969,SolarFlux2015}. Non-helical shear dynamos invoke turbulent resistivity and the shear-current effect~\cite{RogachevskiiKleeorin,Yousef2008,SquireBhattacharjee} to grow the large-scale field. These theories express the emf in terms of the mean magnetic field and current using approximate mean-field closures. They primarily address field growth; the $\alpha\Omega$ equations also produce dynamo waves but rely on a mean-field closure rather than a first principles calculation. Global solar simulations reproduce magnetic reversals, migrating fields, or solar-like differential rotation, but no calculation at solar parameters yet reproduces all measured periods, field phases and strengths, flows, and sunspot migration together~\cite{Ghizaru,Kapyla2012,HottaKusano,Charbonneau2020}.

In accretion disks, Lesur and Ogilvie studied the cyclic regeneration of poloidal field by nonaxisymmetric MRI perturbations and its conversion into toroidal field by shear~\cite{Lesur}. H\'erault et al. located periodic trajectories numerically in three-dimensional magnetohydrodynamic (MHD) simulations; Riols et al. investigated their stability and transitions toward chaos~\cite{Herault,Riols}. Complementary analytical studies explain oscillatory MRI saturation near the instability threshold: Liverts et al. identified a feedback from compressive waves excited by magnetic pressure in stratified disks~\cite{Liverts}, while Vasil derived nonlinear oscillations regulated by the redistribution of mean shear and magnetic flux~\cite{Vasil}. Both expand about the instability threshold with an imposed axial field. Banik et al. explained long toroidal cycles through the interference between shear Alfv\'en waves in quasilinear theory, supported by shearing-box simulations~\cite{Banik}. Here we seek exact analytical solutions with a particular spatial form, retaining the self-consistent non-linear evolution of meridional circulation, shear, and magnetic fields.

Exact nonlinear solutions can be constructed using spatial forms preserved by the MHD equations. Channel modes cancel their perturbation nonlinearities~\cite{Goodman,Lesaffre}. On the other hand, fields linear in position, called affine fields, retain nonlinear advection and magnetic tension because these terms remain linear in position. Holm derived ideal, compressible affine motions that stretch or contract differently along each direction while rotating and transporting magnetic flux, with conserved quantities derived from their symmetries~\cite{Holm}. Craik and Cao constructed further time-dependent affine families~\cite{Craik,Cao}. Arter studied ellipsoidal magnetic dynamics, and Roberts, Shkoller, and Sideris classified recurrent motions of deforming magnetic ellipses~\cite{Arter,Roberts}.

Building on the affine frameworks of Holm, Craik, and Cao, we derive a family of exact solutions to incompressible MHD in background rotational shear. They reveal a cyclic circulation--shear--tension feedback (Fig.~\ref{fig:feedback}) and analytical relations among magnetic-cycle period, field-strength ratio, and circulation speed. Our dynamo wave solution carries finite magnetic torque; Craik's displayed steady-flow reversal has zero magnetic tension. Our meridional velocity and magnetic field can be non-aligned, extending beyond Hamabata's alignment assumption~\cite{Hamabata}. Our fields can carry flux across a local shear region, whereas the moving-boundary fields of Roberts, Shkoller, and Sideris are tangent to an enclosing ellipse~\cite{Roberts}. Two conserved quantities reduce the dynamics to three coupled equations, making the system very nearly integrable and enabling a complete classification of steady-state/fixed point families and analytically tractable non-linear magnetic cycles.

Applied to the Sun's near-surface shear layer (NSSL), the dynamo wave supports stable 11-year reversals at moderate shear, with poleward speeds of $8$--$12\,\mathrm{m\,s^{-1}}$ and poloidal and toroidal amplitudes of approximately $4$--$11$ G and $2.2$--$3.8$ kG near 1 Mm depth, in reasonable agreement with regional observations. These waves also admit a continuous range of slower modulations that includes the observed Gleissberg and Suess--de Vries bands, although the model does not preferentially select their periods. Stronger shear offers a distinct connection: an 88-year dynamo wave supports slow modulation near $203$--$207$ yr without precise shear tuning. This longer wave requires weaker poloidal fields; its magnetic energy varies with an 88-year period and could contribute to long-period activity, although an 88-year polarity reversal remains observationally unestablished. The NSSL more readily satisfies the joint period, field, and flow constraints than the tachocline, where the higher density and inferred return flow require much stronger fields than recent large-scale-field reconstructions. Strong-field tachocline solutions remain possible under seismic upper limits. Our analysis is local and therefore does not address the issue of a global dynamo cycle spanning the convection zone. 

We would like to emphasize that our main goal in this paper is to explain the physics of the magnetic cycle, with the solar comparison serving as a test of physical plausibility rather than a precise fit to the data. A key prediction of our theory is that {\textit {a magnetic cycle many rotation periods long owes its long duration to a large toroidal-to-poloidal field ratio}}. This follows from the fact that when shear acts on a weak poloidal field, it changes only a small fraction of the stronger toroidal field per rotation. A weak poloidal field also limits the magnetic torque, allowing steady balance with advection and Coriolis forces only with a slow meridional circulation. Efficient generation of poloidal flux, as in stratified convection~\cite{Kapyla2012}, would shorten the magnetic cycle. This hints at the solar cycle being primarily driven by shear rather than convection, although a definitive statement would require a detailed analysis of dynamos in the presence of both shear and stratification.

\section{Governing equations}
To construct the solutions, we use radial, azimuthal, and vertical coordinates $x,y,z$ to span a local frame rotating with angular frequency $\Omega>0$, with unit vector $\boldsymbol e_i$ along coordinate $i$; $x$ points outward perpendicular to the rotation axis, $y$ is azimuthal, and $z$ is parallel to the rotation axis. The background azimuthal velocity is $-q\Omega x\boldsymbol e_y$, where $q=-\dd\ln\Omega/\dd\ln R$ is the shear at reference radius $R$. $\boldsymbol U$ is the total velocity in this non-inertial frame. We assume an unstratified medium of constant density $\rho$, i.e., restrict ourselves to length scales smaller than the scale height. Define $\boldsymbol B=\boldsymbol B_{\mathrm{phys}}/\sqrt{4\pi\rho}$, with physical field $\boldsymbol B_{\mathrm{phys}}$, and total pressure per unit density $P=p/\rho+|\boldsymbol B|^2/2$, with gas pressure $p$. The incompressible ideal MHD equations are given by
\begin{align}
&\partial_t\boldsymbol U+\boldsymbol U\cdot\nabla\boldsymbol U
=-\nabla P+\boldsymbol B\cdot\nabla\boldsymbol B\nonumber\\
&\qquad\qquad\qquad\quad\;\,-2\Omega\boldsymbol e_z\times\boldsymbol U+2q\Omega^2x\boldsymbol e_x,\label{momentum}\\
&\partial_t\boldsymbol B = \nabla \times (\boldsymbol{U} \times \boldsymbol{B}),\\
&\nabla\cdot\boldsymbol U=\nabla\cdot\boldsymbol B=0.
\label{MHD}
\end{align}
Magnetic tension, pressure gradients, and Coriolis and centrifugal forces accelerate the plasma. The magnetic field evolves through the ideal electric field $\boldsymbol{E} = -\boldsymbol{U}\times\boldsymbol{B}$ generated by the plasma motion. Constant diffusivity and resistivity, if included, would leave the affine solution unchanged, as shown below.

The analysis is axisymmetric: all fields are independent of $y$. The meridional circulation rate $\omega_{\mathrm c}$, azimuthal velocity gradients $S_x=\partial_xU_y,S_z=\partial_zU_y$, and magnetic gradients $b_{ij}=\partial_jB_i$ depend only on time $t$ and have units of inverse time. We assume the following divergence-free affine form~\cite{Holm}, a reasonable assumption for the large-scale $\boldsymbol{U}$ and $\boldsymbol{B}$ fields:
\begin{equation}
\boxed{\begin{aligned}
\boldsymbol U&=(-\omega_{\mathrm c} z,\ S_x\,x+S_z\,z,\ \omega_{\mathrm c} x),\\
\boldsymbol B&=(b_{xx}x+b_{xz}z,\ b_{yx}x+b_{yz}z,\ b_{xz}x-b_{xx}z).
\end{aligned}}
\label{fields}
\end{equation}
The $x,z$ magnetic components are poloidal; the $y$ component is toroidal. A dot denotes $\dd/\dd t$. Fluid elements circulate meridionally, with $\dot x=-\omega_{\mathrm c}z$ and $\dot z=\omega_{\mathrm c}x$ preserving $x^2+z^2$, while azimuthal velocity adds motion along $y$. Constant spatial derivatives keep every nonlinear term in Eqs.~(\ref{momentum})--(\ref{MHD}) linear in position. Substitution and matching the coefficients of $x$ and $z$ thus give the exact evolution equations for the velocity and magnetic gradients (see Appendix~\ref{app:reduction} for a detailed derivation):
\begin{equation}
\boxed{\begin{aligned}
\dot\omega_{\mathrm c}&=-\Omega S_z,\\
\dot S_x&=b_{yx}b_{xx}+b_{yz}b_{xz}-\omega_{\mathrm c}S_z,\\
\dot S_z&=b_{yx}b_{xz}-b_{yz}b_{xx}+\omega_{\mathrm c}(S_x+2\Omega),\\
\dot b_{xx}&=-2\omega_{\mathrm c}b_{xz},\qquad \dot b_{xz}=2\omega_{\mathrm c}b_{xx},\\
\dot b_{yx}&=S_xb_{xx}+S_zb_{xz}-\omega_{\mathrm c}b_{yz},\\
\dot b_{yz}&=S_xb_{xz}-S_zb_{xx}+\omega_{\mathrm c}b_{yx}.
\end{aligned}}\label{seven}
\end{equation}
The first equation follows by combining the radial and vertical momentum equations to eliminate pressure. Define $b_{\mathrm p}=\sqrt{b_{xx}^2+b_{xz}^2}$, the magnitude of the poloidal field gradient, and let $P_0(t)$ be a spatially constant pressure. Integrating the pressure derivatives yields
\begin{align}
P={}&\tfrac12(b_{\mathrm p}^2+\omega_{\mathrm c}^2+2\Omega S_x+2q\Omega^2)x^2\nonumber\\
&+\Omega S_z\,xz+\tfrac12(b_{\mathrm p}^2+\omega_{\mathrm c}^2)z^2+P_0(t).\label{pressure}
\end{align}
Note that $q$ has dropped out of the equations and only lives in the pressure. Radial and vertical induction give the fourth and fifth equations in Eqs.~(\ref{seven}), rotating $(b_{xx},b_{xz})$ at $2\omega_{\mathrm c}$ while preserving $b_{\mathrm p}$. Azimuthal momentum gives the second and third equations; azimuthal induction gives the last two. The affine fields satisfy both divergence constraints and have zero vector Laplacians, so Eqs.~(\ref{seven}) also solve resistive MHD with constant viscosity $\nu$ and resistivity $\eta$. The linear spatial dependence of the quantities makes the solution local; finite realizations exchange energy and magnetic helicity through the boundaries. The self-consistent emf in the above equations admits many equivalent tensor representations in terms of the mean fields and their derivatives. Therefore, guessing individual dynamo coefficients requires care.

\section{Circulation--shear--tension equations}
\begin{figure*}[t!]
\centering
\includegraphics[width=\textwidth]{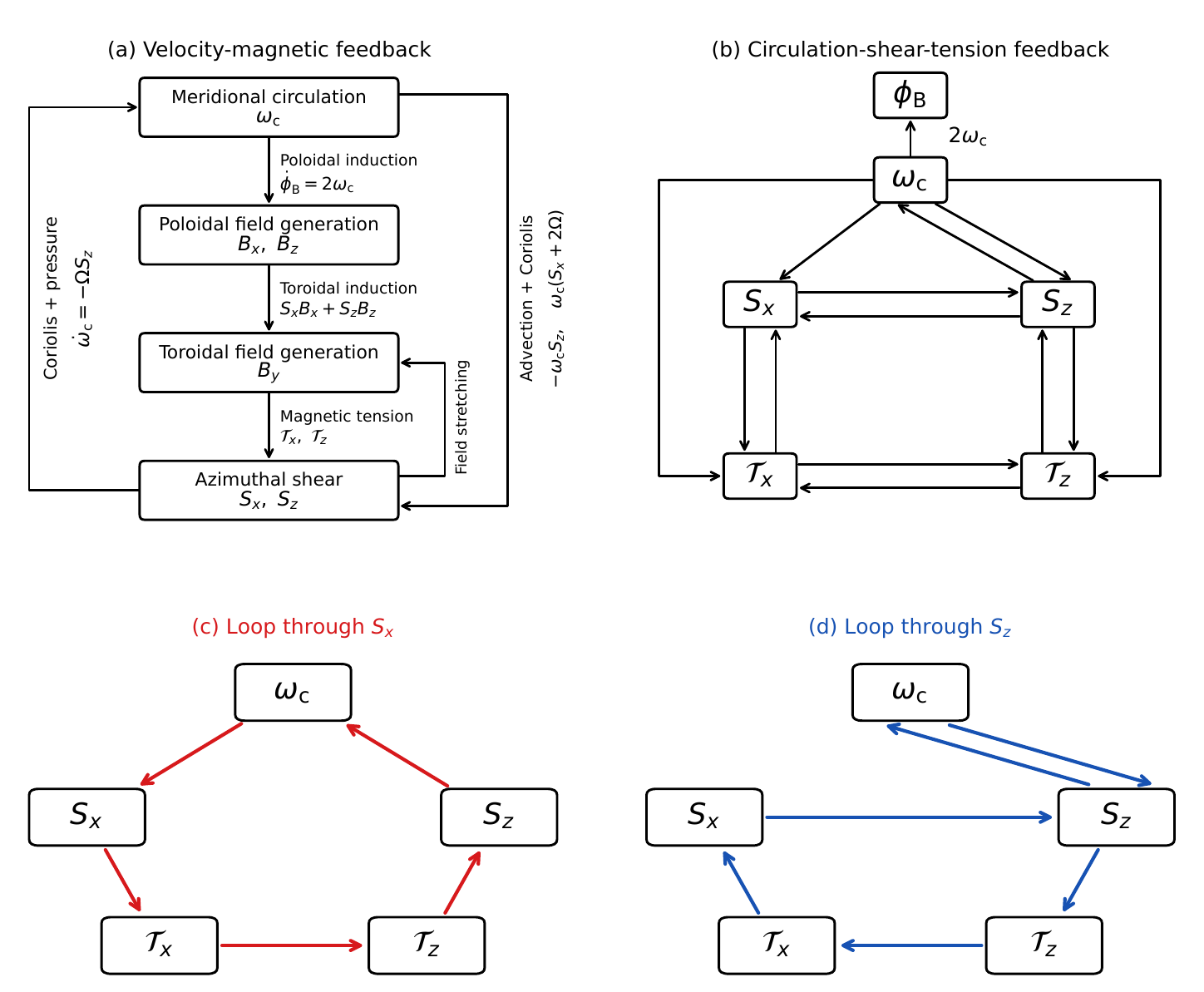}
\caption{The magnetic-cycle feedback. (a) Circulation rotates the poloidal field while preserving $b_{\mathrm p}$; shear generates toroidal field, whose tension changes both shears. Vertical shear changes circulation through Coriolis deflection and pressure. The outer right arrow includes momentum advection and the Coriolis contribution $2\Omega\omega_{\mathrm c}$ to $\dot S_z$. (b) Complete graph of Eqs.~(\ref{five}): arrows connect each source-term variable to the differentiated variable, including both factors of products. (c,d) The red loop includes the $\omega_{\mathrm c}\to S_x$ coupling, while the blue loop bypasses it; both return to circulation through $S_z\to\omega_{\mathrm c}$. Loops through fields or tensions supply dynamo feedback; loops confined to circulation and shear describe hydrodynamic zonal flows and meridional circulation. Circulation advances $\phi_{\mathrm B}$, which does not feed back on the other five variables. Shear acting on a weak poloidal field changes the stronger toroidal field slowly, while the weak magnetic torque balances advection and Coriolis forces only with slow circulation, prolonging the reversal. The dynamo wave maintains constant tension and velocity throughout the reversal.}\label{fig:feedback}
\end{figure*}

We now simplify Eqs.~(\ref{seven}) further, by separating the poloidal-field orientation from circulation, shear, and magnetic tension, which exposes their mutual interplay. Since induction preserves $b_{\mathrm p}$, the poloidal components move in a circle, with their relative angle $\phi_{\mathrm B}$ evolving with time. Define the radial and vertical gradients of azimuthal magnetic tension:
\begin{equation}
\begin{aligned}
(b_{xx},b_{xz})&=b_{\mathrm p}(\cos\phi_{\mathrm B},\sin\phi_{\mathrm B}),\\
\mathcal T_x&=b_{yx}b_{xx}+b_{yz}b_{xz},\\
\mathcal T_z&=b_{yx}b_{xz}-b_{yz}b_{xx}.
\end{aligned}\label{tensions}
\end{equation}
Thus $(\boldsymbol B\cdot\nabla)B_y=\mathcal T_x\,x+\mathcal T_z\,z$; each $\mathcal T_i$ has units of inverse time squared. After differentiating these products with respect to time, Eqs.~(\ref{seven}) reduce to the following circulation--shear--tension equations:
\begin{equation}
\boxed{\begin{aligned}
\dot\omega_{\mathrm c}&=-\Omega S_z,\qquad\qquad\dot\phi_{\mathrm B}&=2\omega_{\mathrm c},\\
\dot S_x&=\mathcal T_x-\omega_{\mathrm c}S_z,\\
\dot S_z&=\mathcal T_z+\omega_{\mathrm c}(S_x+2\Omega),\\
\dot{\mathcal T}_x&=b_{\mathrm p}^2S_x-\omega_{\mathrm c}\mathcal T_z,\\
\dot{\mathcal T}_z&=b_{\mathrm p}^2S_z+\omega_{\mathrm c}\mathcal T_x.
\end{aligned}}\label{five}
\end{equation}
The poloidal field follows from Eqs.~(\ref{tensions}); for $b_{\mathrm p}>0$, the toroidal field is $B_y=(\mathcal T_xB_x+\mathcal T_zB_z)/b_{\mathrm p}^2$. The first four equations above state how vertical shear changes circulation, how circulation advances the poloidal phase, and how circulation and magnetic tension change both shear components. The last two state the reciprocal induction effect: shear induces toroidal-field tension, while circulation rotates its radial and vertical parts. Fig.~\ref{fig:feedback}(b) displays these direct couplings.

For $b_{\mathrm p}>0$, Eqs.~(\ref{five}) are equipped with two integrals of motion. Multiplying the shear and tension equations by their respective variables gives the first integral $C_1$; differentiating $S_z\mathcal T_x-S_x\mathcal T_z$ and using the circulation equation gives the second integral $C_2$ (see Appendix~\ref{app:reduction}):
\begin{equation}
\boxed{\begin{aligned}
C_1&=S_x^2+S_z^2+2\omega_{\mathrm c}^2-
\frac{\mathcal T_x^2+\mathcal T_z^2}{b_{\mathrm p}^2},\\
C_2&=S_z\mathcal T_x-(S_x+2\Omega)\mathcal T_z-2b_{\mathrm p}^2\omega_{\mathrm c}.
\end{aligned}}\label{constants}
\end{equation}
Since $\mathcal T_x^2+\mathcal T_z^2=b_{\mathrm p}^2(b_{yx}^2+b_{yz}^2)$, we can rewrite $C_1$ as
\begin{equation}
C_1-2b_{\mathrm p}^2=|\nabla\boldsymbol U|^2-|\nabla\boldsymbol B|^2,
\label{gradient_energy_constant}
\end{equation}
where each squared gradient sums the squares of its components. This conserved difference is proportional to the kinetic minus magnetic energy in a region with equal averages of $x^2$ and $z^2$ and zero average of $xz$.

$C_2$ can be expressed in terms of the absolute vorticity that includes frame rotation, and the gradient of the current-helicity density $\boldsymbol B\cdot\boldsymbol J$, with current $\boldsymbol J=\nabla\times\boldsymbol B$. Using Eqs.~(\ref{tensions}) and (\ref{fields}) in the second of Eqs.~(\ref{constants}) yields
\begin{align}
C_2&=\boldsymbol\omega_{\rm abs}\cdot\boldsymbol n_{\mathrm B},\nonumber\\\displaybreak[1]
\boldsymbol\omega_{\rm abs}&=\nabla\times\boldsymbol U+2\Omega\boldsymbol e_z,\nonumber\\\displaybreak[1]
\boldsymbol n_{\mathrm B}&=\partial_x\boldsymbol B\times\partial_z\boldsymbol B\nonumber\\\displaybreak[1]
&=b_{\mathrm p}^2\boldsymbol e_y+\boldsymbol e_y\times\nabla(\boldsymbol B\cdot\boldsymbol J)\nonumber\\\displaybreak[1]
&=(-\mathcal T_x,b_{\mathrm p}^2,-\mathcal T_z),\nonumber\\\displaybreak[1]
\boldsymbol B\cdot\boldsymbol J&=\mathcal T_z x-\mathcal T_x z.
\label{current_helicity_normal}
\end{align}
Thus $C_2$ couples absolute vorticity to the poloidal and current-helicity gradients, with $b_{\mathrm p}^2\boldsymbol e_y$ surviving at zero current. Induction and vorticity evolution preserve this product: velocity-gradient contributions cancel, and the curl of magnetic tension, $(-\mathcal T_z,0,\mathcal T_x)$, is perpendicular to $\boldsymbol n_{\mathrm B}$.

These local constants differ from the standard volume integrals of total energy, cross helicity (velocity--magnetic field alignment), and magnetic helicity (vector potential---magnetic field alignment, denoting field linkage and twist). The affine fields grow with position and allow boundary fluxes for these integrals. Our constants also differ from the deformation-energy and rotation constants of magnetic ellipses~\cite{Roberts}. Together, $b_{\mathrm p},C_1$ and $C_2$ constrain the evolution of circulation, shear, and magnetic tension.

For nonzero magnetic tension, these constants determine $S_x$ and $S_z$ once $\omega_{\mathrm c},\mathcal T_x$, and $\mathcal T_z$ are known. To write this algebraic inversion compactly, define $\mathcal T^2=\mathcal T_x^2+\mathcal T_z^2$ and
\begin{equation}
\begin{aligned}
H&=C_2+2b_{\mathrm p}^2\omega_{\mathrm c}+2\Omega\mathcal T_z,\\
\mathcal W&=\pm\sqrt{\mathcal T^2(C_1-2\omega_{\mathrm c}^2+\mathcal T^2/b_{\mathrm p}^2)-H^2}.
\end{aligned}\label{root}
\end{equation}
Here $\mathcal W=S_x\mathcal T_x+S_z\mathcal T_z$, which fixes the sign of the square root from the initial state. Substitution into Eq.~(\ref{five}) leaves three coupled evolution equations:
\begin{equation}
\begin{aligned}
\dot\omega_{\mathrm c}&=-\Omega\frac{\mathcal T_z\mathcal W+\mathcal T_xH}{\mathcal T^2},\\
\dot{\mathcal T}_x&=b_{\mathrm p}^2\frac{\mathcal T_x\mathcal W-\mathcal T_zH}{\mathcal T^2}
-\omega_{\mathrm c}\mathcal T_z,\\
\dot{\mathcal T}_z&=b_{\mathrm p}^2\frac{\mathcal T_z\mathcal W+\mathcal T_xH}{\mathcal T^2}
+\omega_{\mathrm c}\mathcal T_x.
\end{aligned}\label{three}
\end{equation}
These equations apply when tension is nonzero and the sign of $\mathcal W$ is unchanged; otherwise, we should resort to Eqs.~(\ref{five}) for the full solution. In general, though, only three quantities, $\omega_{\mathrm c}, {\mathcal T}_x$ and ${\mathcal T}_z$, determine the full circulation-shear-tension state, with magnetic direction following from $\dot\phi_{\mathrm B}=2\omega_{\mathrm c}$.

\subsection{Why cycles can occur}
The reduced equations (Eqs.~[\ref{seven}], [\ref{five}] or [\ref{three}]) expose the feedback cycle shown in Fig.~\ref{fig:feedback}. Meridional circulation rotates the poloidal field through $\dot\phi_{\mathrm B}=2\omega_{\mathrm c}$. Radial and vertical shear then stretch its components into toroidal field through $S_xB_x+S_zB_z$ in the induction equation. Together, the poloidal and toroidal fields produce azimuthal tension $\mathcal T_x x+\mathcal T_z z$, changing both shears and subsequent toroidal-field production. The changed vertical shear alters the radial Coriolis acceleration; together with pressure gradients, this changes the meridional velocities, giving $\dot\omega_{\mathrm c}=-\Omega S_z$. Circulation then rotates the poloidal field again, completing the loop. Magnetic and kinetic energies are exchanged by induction and the Lorentz force, while the Coriolis force re-orients a velocity component without doing any work.

The cycle can also be understood from the interplay of circulation, shear and tension. Circulation couples the two shears through momentum advection, $-\omega_{\mathrm c}S_z$ in $\dot S_x$ and $+\omega_{\mathrm c}S_x$ in $\dot S_z$; azimuthal Coriolis acceleration adds $2\Omega\omega_{\mathrm c}$ to $\dot S_z$. Induction similarly rotates the tension components through $-\omega_{\mathrm c}\mathcal T_z$ and $+\omega_{\mathrm c}\mathcal T_x$ in Eq.~(\ref{five}). Fig.~\ref{fig:feedback}(a,b) shows this complete feedback, while panels~(c,d) trace two magnetic routes through the radial and vertical shear that reveal the dynamo cycle. Loops confined to circulation and shear describe hydrodynamic zonal flows and meridional circulation; the dynamo routes have to pass through magnetic fields or tensions.

The toroidal-to-poloidal field ratio sets the pace of this feedback relative to rotation. Shear acting on a weak poloidal field changes only a small fraction of the stronger toroidal field per rotation. The poloidal field also transmits the toroidal field’s magnetic tension back to the flow. In this axisymmetric system, that azimuthal tension is $(B_x\partial_x+B_z\partial_z)B_y$, so weakening the poloidal field reduces the torque for a given toroidal-field gradient. For steady vertical shear, $\mathcal T_z=-\omega_{\mathrm c}(S_x+2\Omega)$ balances this torque against momentum advection and Coriolis deflection. At fixed shear and rotation, a weaker torque therefore requires slower circulation. The circulation then turns the poloidal field gradually, reversing the toroidal-induction source over many rotations. The dynamo wave below maintains this balance with constant velocity and tension throughout the magnetic reversal.

Both shear and rotation are essential to this feedback. For $\Omega>0$, maintaining $S_x=S_z=0$ forces $\omega_{\mathrm c}=0$; at $\Omega=0$, circulation is constant and the shear--tension frequencies $\pm\omega_{\mathrm c}\pm i b_{\mathrm p}$ preclude periodic feedback for $b_{\mathrm p}>0$. Without rotation, a steady circulation can still passively rotate an existing poloidal field when azimuthal shear and toroidal field vanish: the Lorentz force is then zero. In all cases, conserved $b_{\mathrm p}$ means that these cycles reverse and redistribute an existing poloidal field. Maintaining it against turbulent diffusion requires regeneration beyond the affine solution.

\subsection{Fixed points and their stability}
We first identify the steady states or fixed points of the circulation--shear--tension Eqs.~(\ref{five}) and their stability under perturbations. A fixed point of Eqs.~(\ref{five}) has constant circulation, shear, and tension, and corresponds to steadily evolving velocity and magnetic fields. For $\Omega>0$, this requires $S_z=\mathcal T_x=0$. The remaining force and induction balances, $\mathcal T_z+\omega_{\mathrm c}(S_x+2\Omega)=0$ and $b_{\mathrm p}^2S_x-\omega_{\mathrm c}\mathcal T_z=0$, give the following fixed point equations:
\begin{equation}
\boxed{\begin{aligned}
S_z&=0,& \mathcal T_x&=0,\\
S_x&=-\frac{2\Omega\omega_{\mathrm c}^2}
{\omega_{\mathrm c}^2+b_{\mathrm p}^2},&
\mathcal T_z&=\frac{b_{\mathrm p}^2}{\omega_{\mathrm c}}S_x .
\end{aligned}}\label{fixed_curve}
\end{equation}
The $\mathcal T_z$ expression requires $\omega_{\mathrm c}\ne0$. For $\omega_{\mathrm c}=0$, the fixed-point equations give $\mathcal T_z=0$, with $S_x=0$ if $b_{\mathrm p}>0$ and arbitrary $S_x$ if $b_{\mathrm p}=0$. At fixed $b_{\mathrm p}$ and $\Omega$, each nonzero circulation selects steady $S_x$ and $\mathcal T_z$.

The different fixed point families are described as follows (see Appendix~\ref{app:stability} for details):
\begin{itemize}
\item \textbf{Steady-circulation reversal (dynamo wave) family.} Here $b_{\mathrm p}>0$ and $\omega_{\mathrm c}\ne0$, with $S_z=\mathcal T_x=0$ and $S_x,\mathcal T_z$ given by Eqs.~(\ref{fixed_curve}). Writing $S_x=-q\Omega$ gives
\begin{equation}
\boxed{\begin{aligned}
\omega_{\mathrm c}^2&=\frac{q}{2-q}b_{\mathrm p}^2,\\
\mathcal T_z&=-(2-q)\Omega\omega_{\mathrm c},
\qquad 0<q<2.
\end{aligned}}
\label{fixed_magnetized}
\end{equation}
The flow is $\boldsymbol U=(-\omega_{\mathrm c}z,-q\Omega x,\omega_{\mathrm c}x)$: meridional circulation (we take $\omega_{\mathrm c}>0$ to specify the circulation direction) and azimuthal shear are both steady. The magnetic-tension force is
$b_{\mathrm p}^2x\boldsymbol e_x+\mathcal T_z z\boldsymbol e_y+b_{\mathrm p}^2z\boldsymbol e_z$. Its radial and vertical parts enter the pressure balance, while $\mathcal T_z z\boldsymbol e_y$ supplies the magnetic torque that maintains the shear. Both poloidal and toroidal fields undergo a periodic reversal. With initial poloidal angle $\phi_{\mathrm B0}$, their evolution is
\begin{equation}
\boxed{\begin{aligned}
\phi_{\mathrm B}&=2\omega_{\mathrm c}t+\phi_{\mathrm B0},\\
B_x&=b_{\mathrm p}(x\cos\phi_{\mathrm B}+z\sin\phi_{\mathrm B}),\\
B_z&=b_{\mathrm p}(x\sin\phi_{\mathrm B}-z\cos\phi_{\mathrm B}),\\
B_y&=-\frac{q\Omega}{\omega_{\mathrm c}}B_z\\
&= \frac{q\Omega b_{\mathrm p}}{\omega_{\mathrm c}}\left(-x\sin\phi_{\mathrm B} + z\cos\phi_{\mathrm B}\right).
\end{aligned}}\label{reversal}
\end{equation}
All three magnetic components are therefore present in general and evolve through induction. The velocity and magnetic tension remain steady, but the magnetic field reverses after $T_{\mathrm B}/2$. Here $T_{\mathrm B}$ is the full signed-field period and $T_{\mathrm{orb}}$ is the rotation period:
\begin{equation}
\frac{T_{\mathrm B}}{T_{\mathrm{orb}}}
=\frac12\sqrt{\frac{2-q}{q}}\frac{\Omega}{b_{\mathrm p}},
\qquad T_{\mathrm{orb}}=\frac{2\pi}{\Omega}.
\label{period}
\end{equation}
Eq.~(\ref{reversal}) gives $\partial_tB_y=-2q\Omega B_x$ at a fixed position, with equal contributions from shear and meridional advection. Thus $\max|\partial_tB_y|/B_{y,\max}=2q\Omega B_{x,\max}/B_{y,\max}=2\omega_{\mathrm c}$, where $B_{i,\max}$ denotes the full-cycle amplitude. At fixed $q$ and $\Omega$, Eq.~(\ref{fixed_magnetized}) gives $\omega_{\mathrm c}=b_{\mathrm p}\sqrt{q/(2-q)}$: induction and circulation both slow down as $b_{\mathrm p}$ decreases, while the toroidal-gradient amplitude $\Omega\sqrt{q(2-q)}$ remains finite.

Let $L_x,L_z$ be the characteristic radial and vertical lengths, $a=L_z/L_x$ their aspect ratio, and $\overline B_{\mathrm p}^2=b_{\mathrm p}^2(L_x^2+L_z^2)/2$ a squared characteristic poloidal amplitude in Alfv\'en-speed units. Then we have
\begin{equation}
\boxed{\frac{T_{\mathrm B}}{T_{\mathrm{orb}}}
=\frac12\sqrt{\frac{2-q}{q}}
\sqrt{\frac{1+a^2}{2}} \frac{\Omega L_x}{\overline B_{\mathrm p}}.}
\label{period_aspect}
\end{equation}
At fixed $b_{\mathrm p}$, the period is independent of aspect ratio. At fixed $\overline B_{\mathrm p}$ and $L_x$, it increases monotonically with $a=L_z/L_x$ (see \cite{Banik} for a similar trend in MRI dynamos).

The stability of this periodic orbit follows by perturbing the five shear--tension variables with time dependence $e^{-i\omega t}$, where $\omega$ is the perturbation frequency and $i^2=-1$. The resulting dispersion relation (between $\omega$ and the physical parameters) is given by
\begin{equation}
\boxed{\begin{aligned}
&\omega\bigl(\omega^4-A_2\omega^2+A_0\bigr)=0,\\
&A_2=(2-q)\Omega^2+
\frac{4(q-1)b_{\mathrm p}^2}{2-q},\\
&A_0=2(2q-1)\Omega^2b_{\mathrm p}^2+
\frac{4b_{\mathrm p}^4}{(2-q)^2}.
\end{aligned}}\label{fixed_magnetized_dispersion}
\end{equation}
The zero-frequency disturbance moves the system to a neighboring
member of the steady family. The other four perturbations form two
frequency pairs. For $0<q<2$, their character is
\begin{subequations}\label{stability}
\begin{gather}
\scalebox{0.9}{$\displaystyle\boxed{\begin{gathered}
\text{two distinct real frequency pairs:}\\[-3pt]
\frac{3-\sqrt5}{4}<q<2,\\
\max\!\left[0,\frac{(1-2q)(2-q)^2}{2}\right]
<\frac{b_{\mathrm p}^2}{\Omega^2}\\[-2pt]
<\frac{2-q}{4}
\left(\sqrt{\frac{2}{q}}-1\right);
\end{gathered}}$}\\\displaybreak[1]
\scalebox{0.9}{$\displaystyle\boxed{\begin{gathered}
\text{one real and one imaginary frequency pair:}\\[-3pt]
0<q<\frac12,\quad
0<\frac{b_{\mathrm p}^2}{\Omega^2}
<\frac{(1-2q)(2-q)^2}{2};
\end{gathered}}$}\\\displaybreak[1]
\scalebox{0.9}{$\displaystyle\boxed{\begin{gathered}
\text{two distinct imaginary frequency pairs:}\\[-3pt]
0<q<\frac{3-\sqrt5}{4},\\
\frac{(1-2q)(2-q)^2}{2}
<\frac{b_{\mathrm p}^2}{\Omega^2}
<\frac{2-q}{4}
\left(\sqrt{\frac{2}{q}}-1\right);
\end{gathered}}$}\\\displaybreak[1]
\scalebox{0.9}{$\displaystyle\boxed{\begin{gathered}
\text{complex-conjugate values of }\omega^2\text{:}\\[-3pt]
0<q<2,\quad
\frac{b_{\mathrm p}^2}{\Omega^2}
>\frac{2-q}{4}
\left(\sqrt{\frac{2}{q}}-1\right).
\end{gathered}}$}
\end{gather}
\end{subequations}
The first case gives bounded oscillations; the second gives one exponentially growing and one decaying disturbance; the third gives two of each. The fourth gives oscillations with growing or decaying amplitudes because the frequencies have both real and imaginary parts. At the lower nonzero boundary, $b_{\mathrm p}^2/\Omega^2=(1-2q)(2-q)^2/2$, which exists for $0<q<1/2$, one frequency pair passes through zero. At the upper boundary, $b_{\mathrm p}^2/\Omega^2=(2-q)(\sqrt{2/q}-1)/4$, the two $\omega^2$ values coincide before becoming complex.

Along this orbit, $\nabla C_1=-2\omega_{\mathrm c}\nabla C_2/b_{\mathrm p}^2$, with gradients taken over the five shear--tension variables. Therefore, the conservation laws impose only one independent linear condition, allowing four nonzero frequencies about the dynamo wave fixed point despite a three-variable reduction (Eqs.~[\ref{three}]) of the general motion.

\item \textbf{Stationary poloidal family:} $b_{\mathrm p}>0$ and $\omega_{\mathrm c}=0$, with
\begin{equation}
S_x=S_z=\mathcal T_x=\mathcal T_z=0.
\end{equation}
The velocity and shear vanish: $\boldsymbol U=0$. The azimuthal tension and magnetic torque vanish, while the radial--vertical tension $b_{\mathrm p}^2(x\boldsymbol e_x+z\boldsymbol e_z)$ enters the pressure balance together with the tidal force. The field is stationary and purely poloidal:
\begin{equation}
\begin{aligned}
B_x&=b_{\mathrm p}(x\cos\phi_{\mathrm B0}+z\sin\phi_{\mathrm B0}),\\
B_y&=0,\\
B_z&=b_{\mathrm p}(x\sin\phi_{\mathrm B0}-z\cos\phi_{\mathrm B0}),
\end{aligned}
\label{stationary_poloidal_field}
\end{equation}
where the constant angle $\phi_{\mathrm B0}$ is arbitrary. Linearizing the seven coefficients in Eq.~(\ref{seven}) gives the dispersion relation
\begin{equation}
\omega^3(\omega^2+b_{\mathrm p}^2)
(\omega^2+b_{\mathrm p}^2-2\Omega^2)=0.
\label{fixed_poloidal}
\end{equation}
Every member is unstable because $\omega=\pm i b_{\mathrm p}$ gives exponential growth. The second pair, $\omega=\pm\sqrt{2\Omega^2-b_{\mathrm p}^2}$, oscillates when $b_{\mathrm p}<\sqrt2\Omega$, reaches zero frequency at equality, and contains a growing disturbance when $b_{\mathrm p}>\sqrt2\Omega$.

\item \textbf{Toroidal-rotation family:} $b_{\mathrm p}=0$ and $\omega_{\mathrm c}\neq0$, with
\begin{equation}
S_x=-2\Omega,\qquad
S_z=\mathcal T_x=\mathcal T_z=0.
\end{equation}
The flow $\boldsymbol U=(-\omega_{\mathrm c}z,-2\Omega x,\omega_{\mathrm c}x)$ has steady meridional circulation and azimuthal shear. The magnetic field is purely toroidal and independent of $y$, so its tension and magnetic torque vanish. For zero toroidal amplitude, the dispersion relation is
\begin{equation}
\omega(\omega^2-\omega_{\mathrm c}^2)^2
(\omega^2-4\omega_{\mathrm c}^2)=0.
\label{fixed_center}
\end{equation}
Every nonzero-frequency disturbance is bounded; the zero-frequency disturbance changes the constant circulation and moves the state to a neighboring member of the family. The velocity disturbances obey
\begin{equation}
\begin{aligned}
\delta\dot\omega_{\mathrm c}+\Omega\delta S_z&=0,\\
\delta\dot S_x+\omega_{\mathrm c}\delta S_z&=0,\\
\delta\dot S_z-\omega_{\mathrm c}\delta S_x&=0.
\end{aligned}\label{hydro_linear}
\end{equation}
The last two equations give
$\delta\ddot S_x+\omega_{\mathrm c}^2\delta S_x=0$
and the same oscillator equation for $\delta S_z$.
Both shear disturbances therefore oscillate at $\omega_{\mathrm c}$,
as does the circulation disturbance through the first equation.
Poloidal induction gives
$\delta\dot b_{xx}=-2\omega_{\mathrm c}\delta b_{xz}$ and
$\delta\dot b_{xz}=2\omega_{\mathrm c}\delta b_{xx}$,
producing magnetic oscillations at $2\omega_{\mathrm c}$ with
toroidal-to-poloidal gradient ratio $2\Omega/\omega_{\mathrm c}$.
Slow circulation therefore gives a long magnetic period and a large
toroidal component.

These are the leading magnetic oscillations of the nearby dynamo wave.
As $b_{\mathrm p}\to0$ at fixed nonzero $\omega_{\mathrm c}$, its
shear $S_x=-2\Omega\omega_{\mathrm c}^2/(\omega_{\mathrm c}^2+b_{\mathrm p}^2)$
approaches $-2\Omega$, its tension
$\mathcal T_z=b_{\mathrm p}^2S_x/\omega_{\mathrm c}$ vanishes,
and both magnetic components tend to zero, recovering the member
analyzed above.

For finite toroidal gradient amplitude $b_{\mathrm t}$ and initial
phase $\psi_0$, the exact magnetic evolution is
\begin{equation}
\begin{aligned}
b_{yx}&=b_{\mathrm t}\cos(\omega_{\mathrm c}t+\psi_0),\\
b_{yz}&=b_{\mathrm t}\sin(\omega_{\mathrm c}t+\psi_0),
\end{aligned}\label{toroidal_rotation}
\end{equation}
giving
$B_y=b_{\mathrm t}[x\cos(\omega_{\mathrm c}t+\psi_0)
+z\sin(\omega_{\mathrm c}t+\psi_0)]$. The background toroidal field rotates at $\omega_{\mathrm c}$, while the poloidal disturbance rotates at $2\omega_{\mathrm c}$. Their coupling
produces magnetic tension at the difference frequency $\omega_{\mathrm c}$,
which is also the natural frequency of the velocity disturbances.
The flow is therefore driven resonantly: each forcing cycle reinforces
the velocity oscillation, making its amplitude undergo a secular growth linearly in time.

\item \textbf{Stationary shear--toroidal family:} $b_{\mathrm p}=\omega_{\mathrm c}=0$, with
\begin{equation}
\begin{aligned}
S_z&=\mathcal T_x=\mathcal T_z=0,
&S_x&\ \hbox{arbitrary},\\
B_x&=B_z=0,
&B_y&=b_{yx}x+b_{yz}z,
\end{aligned}
\end{equation}
where $b_{yx}$ and $b_{yz}$ are arbitrary constants.
The velocity is a steady zonal shear
$\boldsymbol U=(0,S_x\,x,0)$, directed azimuthally.
The stationary toroidal field is independent of $y$, so its tension
and magnetic torque vanish; its magnetic pressure is included in $P$.
This family therefore describes a hydrodynamic equilibrium flow
carrying an optional passive toroidal field.

The dispersion relation is
\begin{equation}
\omega^5[\omega^2-\Omega(S_x+2\Omega)]=0.
\label{fixed_shear}
\end{equation}
Circulation perturbations around $\omega_{\mathrm c}=0$ oscillate
for $S_x>-2\Omega$ and grow exponentially for $S_x<-2\Omega$.
Magnetic disturbances in nonzero shear can grow linearly in time;
at $S_x=-2\Omega$, a finite equilibrium toroidal field can produce
growth up to cubic order in time. For zero shear and zero
toroidal-field amplitude, the zero-frequency disturbances remain constant.
\end{itemize}

The coupling between magnetic tension and rotational shear that governs these non-linear orbits and their (affine) modulations also drives the linear axisymmetric MRI in sinusoidal perturbations. Let us therefore compare the dispersion relation (Eq.~[\ref{fixed_magnetized_dispersion}]) for perturbations around our dynamo wave family to the MRI dispersion relation. Let $\boldsymbol k=(k_x,0,k_z)$ be the wavevector of an axisymmetric sinusoidal disturbance $\sim e^{i\boldsymbol{k}\cdot\boldsymbol{r}}$ in the presence of a uniform vertical field $\boldsymbol{B}_{\mathrm{phys}}$. Denote the Alfv\'en frequency by $\omega_{\mathrm A}=\boldsymbol k\cdot\boldsymbol v_{\mathrm A}$, where $\boldsymbol v_{\mathrm A}=\boldsymbol B_{\mathrm{phys}}/\sqrt{4\pi\rho}$, and define $\mu=k_z^2/(k_x^2+k_z^2)$. For this comparison, we identify $k v_{\mathrm A}=b_{\mathrm p}$, where $k=|\boldsymbol k|$, so that $\omega_{\mathrm A}^2=\mu b_{\mathrm p}^2$. This matches the characteristic Alfv\'en rates of the two spatial forms. The conventional MRI dispersion relation is~\cite{BalbusHawley,BalbusHawleyReview}
\begin{equation}
\begin{aligned}
\omega^4-2\mu\bigl[(2-q)\Omega^2+b_{\mathrm p}^2\bigr]\omega^2+\mu^2b_{\mathrm p}^2
\bigl(b_{\mathrm p}^2-2q\Omega^2\bigr)=0.
\end{aligned}
\label{mri}
\end{equation}
At $\mu=1/2$, its rotation-only coefficient agrees with that of Eq.~(\ref{fixed_magnetized_dispersion}), while the magnetic terms differ. For $1/2\le q<2$ and $\mu>0$, the stability conditions for the perturbations $(\omega^2>0)$ are
\begin{equation}
\boxed{\begin{aligned}
\text{Dynamo wave:}\quad&
0<\frac{b_{\mathrm p}^2}{\Omega^2}
<\frac{2-q}{4}\left(\sqrt{\frac{2}{q}}-1\right),\\
\text{MRI modes:}\quad&
\frac{b_{\mathrm p}^2}{\Omega^2}>2q.
\end{aligned}}
\label{wave_mri_stability}
\end{equation}
Stronger magnetic tension therefore stabilizes the MRI modes, whereas the dynamo wave loses stability at its upper boundary through oscillatory growth. These criteria describe perturbations of different states: sinusoidal disturbances of a uniform field for the MRI, and affine perturbations around a finite-amplitude magnetic cycle for the dynamo wave. Nevertheless, a key difference should be noted: our dynamo waves are stable for weaker poloidal fields, precisely in the regime where the linear axisymmetric MRI modes are unstable. This is probably why MRI dynamo simulations always show the onset of dynamo cycles following an initial MRI, provided the poloidal field remains weaker than the toroidal field, as in the high-plasma-beta, zero-net-flux case \cite{Shi,Banik}.

\subsection{Other exact non-linear orbits}
Beyond the steady state families of orbits identified above, we may have other exactly integrable non-linear orbits. Three choices reduce the five equations to one analytically solvable equation. Appendix~\ref{app:orbits} derives these non-linear orbits by fixing circulation, setting selected physical quantities to zero, or making radial shear depend only on circulation.

First, the \textit{oscillating-circulation cycle} has evolving circulation, shear, and tension. Exact balance requires $\mathcal T_x=\omega_{\mathrm c}S_z/3$ and \(\mathcal T_z\) determined entirely by \(\omega_{\mathrm c}\) with no additional constant term. The resulting solution is an anharmonic oscillator:
\begin{equation}
\begin{aligned}
\ddot\omega_{\mathrm c}&+2b_{\mathrm p}^2\omega_{\mathrm c}
+\tfrac29\omega_{\mathrm c}^3=0,\\
S_x&=\frac{3b_{\mathrm p}^2}{\Omega}-2\Omega
+\frac{\omega_{\mathrm c}^2}{3\Omega},&
S_z&=-\frac{\dot\omega_{\mathrm c}}{\Omega},\\
\mathcal T_x&=-\frac{\omega_{\mathrm c}\dot\omega_{\mathrm c}}{3\Omega},&
\mathcal T_z&=-\frac{b_{\mathrm p}^2\omega_{\mathrm c}}{\Omega}
-\frac{\omega_{\mathrm c}^3}{9\Omega}.
\end{aligned}\label{elliptic}
\end{equation}
For $0<b_{\mathrm p}<\sqrt{2/3}\Omega$, let $A$ be the maximum circulation rate, $\kappa$ its frequency scale, and $m$ the elliptic parameter. The solution is
\begin{equation}
\begin{aligned}
\omega_{\mathrm c}&=A\operatorname{cn}(\kappa t\mid m),\\
A^2&=3b_{\mathrm p}(\sqrt6\Omega-3b_{\mathrm p}),\\
\kappa^2&=2\sqrt6\Omega b_{\mathrm p}/3,\qquad
m=\tfrac12-\sqrt6b_{\mathrm p}/(4\Omega),\\
\phi_{\mathrm B}&=\phi_{\mathrm B0}
+6\arcsin[\sqrt m\operatorname{sn}(\kappa t\mid m)] .
\end{aligned}\label{elliptic_solution}
\end{equation}
The Jacobi functions $\operatorname{cn},\operatorname{sn}$ play the roles of cosine and sine for this anharmonic oscillator, whose oscillation frequency changes with amplitude. The period is $T_{\rm osc}=4K(m)/\kappa$, where \(K(m)=\int_0^{\pi/2}{\mathrm d}\vartheta/\sqrt{1-m\sin^2\vartheta}\) is the complete elliptic integral of the first kind, with integration angle $\vartheta$. Substituting Eq.~(\ref{elliptic}) into Eq.~(\ref{fields}) gives the corresponding periodic velocity and magnetic fields:
\begin{equation}
\begin{aligned}
\boldsymbol U={}&\left(-\omega_{\mathrm c}z,
\left[\frac{3b_{\mathrm p}^2}{\Omega}-2\Omega
+\frac{\omega_{\mathrm c}^2}{3\Omega}\right]x
-\frac{\dot\omega_{\mathrm c}}{\Omega}z,
\omega_{\mathrm c}x\right),\\
B_x={}&b_{\mathrm p}(x\cos\phi_{\mathrm B}+z\sin\phi_{\mathrm B}),\\
B_z={}&b_{\mathrm p}(x\sin\phi_{\mathrm B}-z\cos\phi_{\mathrm B}),\\
B_y={}&-\frac{\omega_{\mathrm c}\dot\omega_{\mathrm c}}
{3\Omega b_{\mathrm p}^2}B_x
-\left(\frac{\omega_{\mathrm c}}{\Omega}
+\frac{\omega_{\mathrm c}^3}{9\Omega b_{\mathrm p}^2}\right)B_z .
\end{aligned}
\label{elliptic_fields}
\end{equation}
The meridional velocity reverses with $\omega_{\mathrm c}$, while the poloidal field rotates and the toroidal field reverses periodically. Because the radial shear depends on $\omega_{\mathrm c}^2$, it completes two oscillations during each circulation cycle; the vertical shear follows $\dot\omega_{\mathrm c}$.

Second, setting $\omega_{\mathrm c}=S_z=\mathcal T_z=0$ gives $\dot S_x=\mathcal T_x$ and $\dot{\mathcal T}_x=b_{\mathrm p}^2S_x$, hence $\ddot S_x-b_{\mathrm p}^2S_x=0$. With $S_1=S_x(0)$ and $b_{\mathrm p}S_2=\mathcal T_x(0)$ setting the initial shear and tension,
\begin{equation}
\begin{aligned}
S_x&=S_1\cosh(b_{\mathrm p}t)+S_2\sinh(b_{\mathrm p}t),\\
\mathcal T_x&=b_{\mathrm p}
[S_1\sinh(b_{\mathrm p}t)+S_2\cosh(b_{\mathrm p}t)] .
\end{aligned}\label{hyperbolic_orbit}
\end{equation}
The poloidal field stays fixed, while radial shear stretches it into a toroidal field (determined by $\mathcal T_x$). The azimuthal velocity and toroidal field admit exponentially growing modes, so this family is unstable. Its stationary limit belongs to the stationary poloidal family.

Third, $b_{\mathrm p}=\mathcal T_x=\mathcal T_z=0$ gives a hydrodynamic family:
\begin{equation}
\begin{aligned}
S_x&=S_{\mathrm h}+\frac{\omega_{\mathrm c}^2}{2\Omega},
\qquad S_z=-\frac{\dot\omega_{\mathrm c}}{\Omega},\\
\ddot\omega_{\mathrm c}
&+\Omega(S_{\mathrm h}+2\Omega)\omega_{\mathrm c}
+\tfrac12\omega_{\mathrm c}^3=0 .
\end{aligned}\label{hydro_orbits}
\end{equation}
Here $S_{\mathrm h}$ is a constant shear. We call this family hydrodynamic because it has no poloidal field or magnetic tension; an optional purely toroidal, axisymmetric field is carried passively. The oscillator couples radial--vertical circulation to azimuthal zonal flow, with circulation reversing or retaining its sign according to the orbit. It describes cycles around the toroidal-rotation centers and the stationary shear--toroidal saddle. Nearby cycles do not separate exponentially, but their periods differ slightly, so they drift out of phase linearly in time. The zero-field equilibrium obeys the usual pressure--Coriolis balance of disk zonal flows~\cite{JohansenZonal,Vanon}.

\section{Solar application}
Let us now examine whether these non-linear magnetic cycles can account for the solar observations. Two solar regions contain strong rotational shear and can therefore participate in an active shear-driven dynamo: the near-surface shear layer (NSSL) in the outer convection zone, where angular velocity varies rapidly with radius, and the tachocline near its base, separating the differentially rotating convection zone from the nearly uniformly rotating radiative interior. Both can stretch a poloidal field into a strong toroidal field. We apply the dynamo wave solution to each region, comparing its period, meridional flow, and magnetic-field strengths with observations.

Sunspot groups reverse their predominant polarity ordering roughly every $11$ years~\cite{HalePolarity1924,ZhukovaHale2022}. Surface observations show that the large-scale radial (poloidal) and azimuthal (toroidal) fields share an approximately 22-year magnetic cycle, with polarity reversals on the 11-year timescale and latitude-dependent phase lags~\cite{CameronFields2018}. A reconstruction from helioseismic rotation measurements likewise finds a dominant 22-year period in both poloidal and toroidal components, assuming that magnetic stresses drive the observed angular-velocity variations~\cite{AntiaChitreGough2013}. Their common period supports our dynamo-wave picture: at a fixed spatial point, both poloidal and toroidal fields oscillate at $2\omega_{\mathrm c}$. This motivates us to suggest that our dynamo wave solution {\textit{is}} the solar cycle.

\subsection{Dynamo wave: period, velocity and field amplitudes}
For the solar comparison, we retain the cylindrical axes of Eq.~(\ref{momentum}): $(x,y)$ span the plane perpendicular to the rotation axis, and $z$ points along the rotation axis. At latitude $\theta$, the local radial and northward directions are $\boldsymbol e_r=\cos\theta\,\boldsymbol e_x+\sin\theta\,\boldsymbol e_z$ and $\boldsymbol e_{\mathrm{lat}}=-\sin\theta\,\boldsymbol e_x+\cos\theta\,\boldsymbol e_z$. We neglect curvature over distances small compared with the solar radius and the distance to the rotation axis, applying this local comparison at low latitudes.

A displaced circulation center permits nonzero local flow. Defining $X=x+x_0$ and $Z=z+z_0$, with constant offsets $x_0,z_0$, gives $\boldsymbol U=(-\omega_{\mathrm c}Z,-q\Omega x,\omega_{\mathrm c}X)$. Translating the magnetic field identically and adding linear pressure terms preserves the solution. At the sampling point, $X=V\cos\theta/\omega_{\mathrm c}$ and $Z=V\sin\theta/\omega_{\mathrm c}$ give a purely latitudinal flow with signed speed $V$. The steady dynamo wave obeys
\begin{equation}
\boxed{\begin{aligned}
\dot\phi_{\mathrm B}&=2\omega_{\mathrm c},
&S_x&=-q\Omega,\\
b_{\mathrm p}^2&=\frac{2-q}{q}\omega_{\mathrm c}^2,
&T_{\mathrm B}&=\frac{\pi}{\omega_{\mathrm c}},\\
B_y&=-\frac{q\Omega}{\omega_{\mathrm c}}B_z.&&
\end{aligned}}
\label{solar_orbit_relations}
\end{equation}
Projecting onto the local radial and latitudinal directions gives the physical field amplitudes:
\begin{equation}
\begin{gathered}
\frac{T_{\mathrm B}}2=\frac{\pi}{2\omega_{\mathrm c}},\\
B_{r,\max}=B_{\mathrm{lat},\max}
=|V|\sqrt{4\pi\rho}\sqrt{\frac{2-q}{q}},\\
\frac{B_{\phi,\max}}{B_{\mathrm{lat},\max}}
=\frac{q\Omega}{\omega_{\mathrm c}}.
\end{gathered}
\label{solar_circular_local_scaling}
\end{equation}
Here $V$ is in $\mathrm{cm\,s^{-1}}$, and each field amplitude is its maximum magnitude at the same point over the magnetic cycle period $T_{\mathrm B}$. The two poloidal components have equal amplitudes and are a quarter-cycle apart because the field rotates in the meridional plane. Eliminating $\omega_{\mathrm c}$ relates the reversal period directly to the toroidal-to-poloidal ratio:
\begin{equation}
\boxed{\frac{T_{\mathrm B}}2
=\frac{T_{\mathrm{orb}}}{4q}
\frac{B_{\phi,\max}}{B_{\mathrm{lat},\max}}.}
\label{period_field_ratio}
\end{equation}
Shear naturally produces a large $B_{\phi,\max}/B_{\mathrm{lat},\max}$  by stretching poloidal into toroidal field~\cite{Shi,Banik}. Stratified convection can instead regenerate substantial poloidal field: K\"apyl\"a et al.~\cite{Kapyla2012} found comparable mean radial and toroidal fields, with stronger radial field at high latitude. Such field ratios would give reversals on the rotation timescale, for $q$ of order unity. The long period ($11$ yr) of the solar cycle relative to the rotation period of $\sim 26$ days therefore calls for shear as the primary actor behind the cycle; convection/buoyancy mainly transport and replenish the poloidal flux.

The dynamo wave’s steady meridional flow gives a fluid circuit time $2\pi/\omega_{\mathrm c}=2T_{\mathrm B}$, or 44 yr for an 11-year reversal. Helioseismic flow reconstructions suggest one cell per hemisphere with an estimated 22-year turnover~\cite{Gizon,GizonTurnover2020}. Cross-equatorial NSSL cells, with opposite return flows below about 20 Mm~\cite{SenCrossEquatorial2026}, could connect the hemispheres without net equatorial mass transport, but whether a 44-year circuit is amenable remains to be seen.

\subsection{Near-surface shear layer and the 11-year cycle}
We first apply these relations to the NSSL. Its rotation rate and radial shear vary with depth and latitude: recent ring-diagram measurements cover 1--17 Mm and resolve an enhanced-shear region with a weaker, shallower flank~\cite{Rabello,RabelloRotation2026,RabelloShear2026}. The measured radial shear is $q_r=-r\partial_r\ln\Omega$, where $r$ is the distance from the Sun's center. The model uses $q=-R\partial_R\ln\Omega$ at fixed cylindrical height, with $R=r\cos\theta$. The chain rule gives $R\partial_R=r\cos^2\theta\,\partial_r-\sin\theta\cos\theta\,\partial_\theta$, and hence
\begin{equation}
q=\cos^2\theta\,q_r+
\sin\theta\cos\theta\,\partial_\theta\ln\Omega.
\label{solar_local_shear_conversion}
\end{equation}
Measurements give $q_r\simeq0.5$--$1$ in moderately sheared portions of the NSSL, rising to $2$--$3$ in the narrow enhanced-shear layer and falling toward zero near its shallow boundary~\cite{Rabello,RabelloShear2026}. Between latitudes $15^\circ$ and $30^\circ$, the shallow rotation profile gives $\Omega/(2\pi)\simeq430$--$450$ nHz and $\partial_\theta\ln\Omega\simeq-0.07$ to $-0.19$, with $\theta$ in radians~\cite{RabelloRotation2026}. Eq.~(\ref{solar_local_shear_conversion}) then gives regional ranges of roughly $q\simeq0.3$--$0.9$ and $1.4$--$2.8$, respectively.

\begin{table*}[t!]
\centering
\caption{Joint constraints on an 11-year NSSL dynamo wave at $\Omega/(2\pi)=443$ nHz, using $B_{\mathrm{lat},\max}=3$--$7$ G, $B_{\phi,\max}=1$--$2$ kG, and $|V|=8$--$12\,\mathrm{m\,s^{-1}}$.}
\label{tab:nssl_constraints}
\renewcommand{\arraystretch}{1.25}
\begin{tabular}{|p{0.20\textwidth}|p{0.37\textwidth}|p{0.33\textwidth}|}
\hline
\textbf{Constraint} & \textbf{Input} & \textbf{Result} \\
\hline
Shear
& Field ratio $\simeq140$--$670$
& $q\simeq0.23$--$1.1$ \\
\hline
Stable shear
& Eq.~(\ref{stability}) for the 11-year wave
& $0.5\lesssim q\lesssim1.1$ within the field-ratio range \\
\hline
Density
& Field and flow amplitudes, jointly restricted to stable shear
& $\rho\simeq1.8\times10^{-7}$--$1.8\times10^{-6}\,\mathrm{g\,cm^{-3}}$ \\
\hline
Depth
& Model S density profile~\cite{ModelS}
& Down to $\sim0.9$ Mm below the photosphere \\
\hline
\end{tabular}
\end{table*}

The remaining constraints come from flow and field measurements. Inversions give shallow poleward speeds of $8$--$12\,\mathrm{m\,s^{-1}}$ at low to middle latitudes, increasing to $15$--$19\,\mathrm{m\,s^{-1}}$ near 14 Mm~\cite{BasuAntiaFlow2010,HerczegJackiewicz2023}. Zonal median radial fields near $\pm80^\circ$ were $7.5$--$10$ G at the 1995--1996 minimum~\cite{ZiegerPolar2019}. Attributing the observed angular acceleration to magnetic stresses, Antia et al.'s Fig.~4~\cite{AntiaChitreGough2013} gives full-record amplitudes of $3$--$7$ G latitudinally and $1$--$2$ kG toroidally near $30^\circ$ and 14 Mm. Seismic fits give toroidal peaks of $0.38$ and $1.4$ kG at 0.7 and 2.8 Mm, respectively, and a surface-normalized dipole of $124\pm17$ G~\cite{Baldner}. These estimates can include unresolved correlations and tangled fields; the dipolar normalization describes a global field geometry. The predicted field strengths also depend on the local density, for which Model S gives $\rho=2.1\times10^{-6}\,\mathrm{g\,cm^{-3}}$ at 1 Mm depth~\cite{ModelS}.

\begin{figure}[t!]
\includegraphics[width=0.48\textwidth]{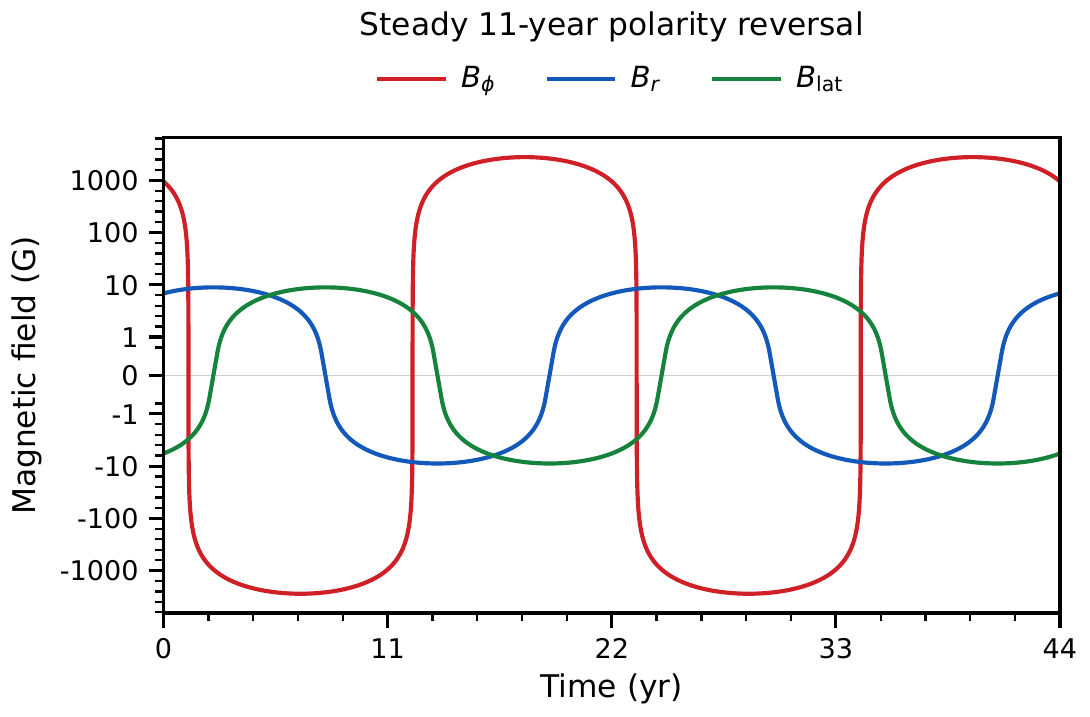}
\caption{The steady dynamo wave at $\Omega/(2\pi)=443$ nHz, $\theta=20^\circ$, $q=0.506$, $\rho=2.1\times10^{-6}\,\mathrm{g\,cm^{-3}}$, and local speed $10\,\mathrm{m\,s^{-1}}$. Toroidal (red), radial (blue), and latitudinal (green) fields reverse every 11 yr. The symmetric logarithmic scale is linear within $\pm0.5$ G.}\label{fig:cycles}
\end{figure}

\begin{figure*}[p]
\centering
\includegraphics[width=\textwidth]{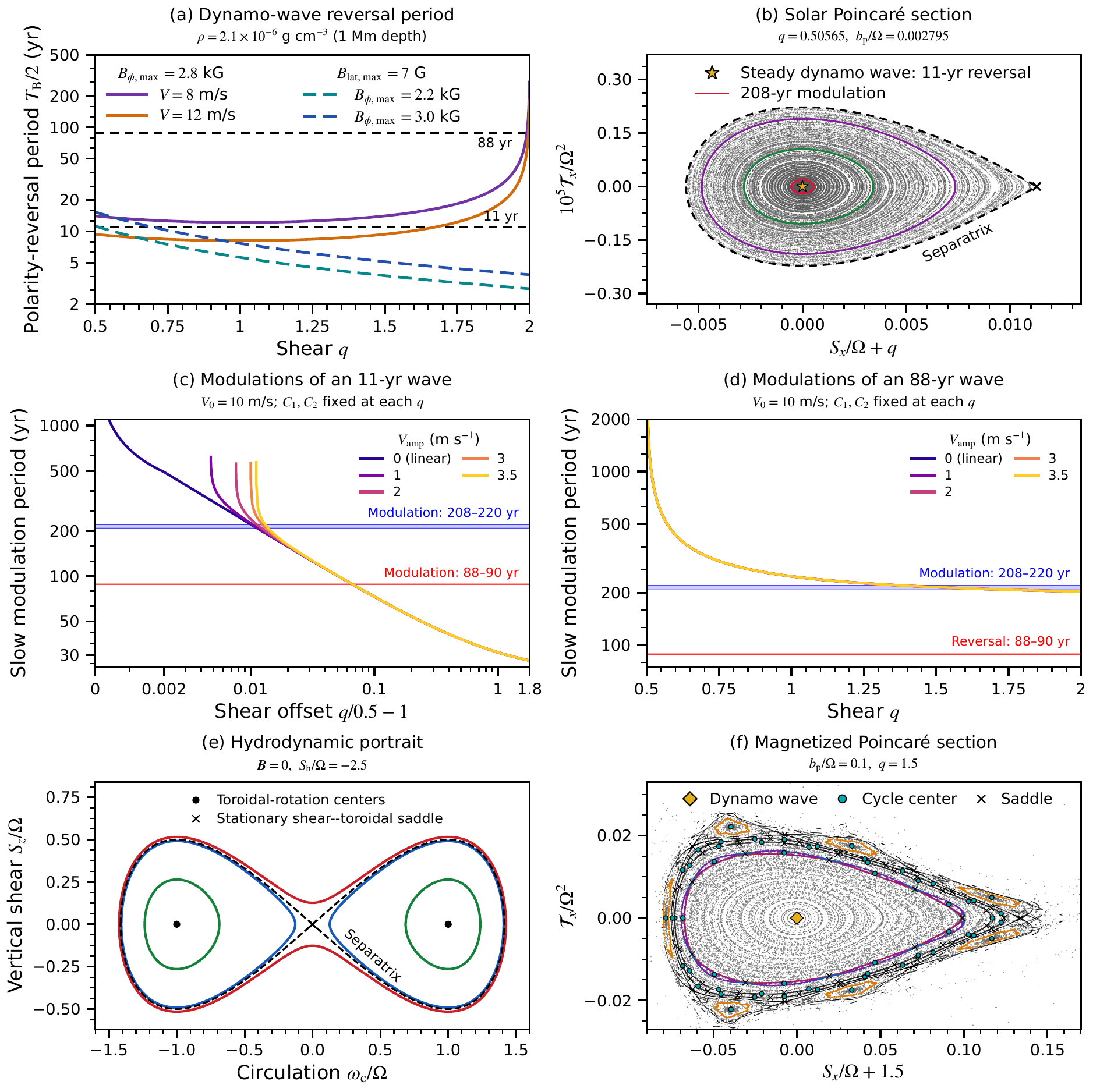}
\caption{Magnetic reversals and their slower modulations.
Solar panels (a--d) use $\Omega/(2\pi)=443$ nHz.
(a) Reversal period at the 1-Mm density
$\rho=2.1\times10^{-6}\,\mathrm{g\,cm^{-3}}$.
Solid curves fix $B_{\phi,\max}=2.8$ kG and $V=8$ or
$12\,\mathrm{m\,s^{-1}}$; dashed curves fix
$B_{\mathrm{lat},\max}=7$ G and $B_{\phi,\max}=2.2$ or
$3.0$ kG. The remaining amplitude follows from
Eq.~(\ref{solar_circular_local_scaling}); black dashed
lines mark 11 and 88 yr.
(b) Solar Poincar\'e section at $q=0.50565$, with
$(b_{\mathrm p}/\Omega,C_1/\Omega^2,C_2/\Omega^3)
=(0.002795,-0.499931,0.00363029)$.
The gold star marks the steady 11-year wave, the crimson
curve a 208-year modulation, and the black cross and
dashed curve an unstable cycle and its separatrix.
(c,d) Slow modulation periods about 11- and 88-year
waves, respectively, with $V_0=10\,\mathrm{m\,s^{-1}}$.
Each $q$ fixes the steady wave's $b_{\mathrm p},C_1,C_2$;
bounded trajectories start at $S_z=\mathcal T_x=0$.
Colors denote $V_{\rm amp}=(V_{\max}-V_{\min})/2$, with
zero denoting the linear limit. Panel (c) uses a logarithmic
shear offset, linear below $q=0.501$.
The red band denotes 88--90-year modulation in (c) and
reversal in (d); blue denotes 208--220-year modulation.
(e) Hydrodynamic motion at $S_{\mathrm h}/\Omega=-2.5$:
the separatrix divides circulation that preserves or
reverses its direction.
(f) Non-solar magnetic section at $q=1.5$, with
$(b_{\mathrm p}/\Omega,C_1/\Omega^2,C_2/\Omega^3)
=(0.1,1.56,0.0398)$.
The gold diamond marks the steady wave; cyan centers
and black saddles mark stable and unstable cycles.
Both sections sample $S_z=0$ with $\dot S_z>0$.
Shear controls the reversal and modulation periods, with an 88-year wave approaching a 203-year slow modulation at strong shear.}
\label{fig:phase}
\end{figure*}

In Fig.~\ref{fig:phase}(a) we plot the dynamo wave periods obtained from representative field and flow amplitudes at this density and $\Omega/(2\pi)=443$ nHz. The solid curves fix $B_{\phi,\max}=2.8$ kG and $V=8$ or $12\,\mathrm{m\,s^{-1}}$, allowing the poloidal amplitude to follow from the shear. Their reversal periods are of order a decade at moderate shear, reach a minimum at $q=1$, and increase toward $q=2$ as the poloidal amplitude decreases. The dashed curves instead fix $B_{\mathrm{lat},\max}=7$ G and $B_{\phi,\max}=2.2$ or $3.0$ kG, with the speed determined by $q$. Their periods decrease as $1/q$, crossing 11 yr at $q\simeq0.51$ and $0.70$, where the implied speeds are approximately $8$ and $10\,\mathrm{m\,s^{-1}}$. Thus field and flow amplitudes close to the regional estimates can produce the observed $11$ yr reversal period at moderate shear.

The same plot also admits much longer reversals. For the fixed-flow curves, $q\simeq1.988$--$1.997$ gives periods of $80$--$100$ yr, with poloidal amplitudes of approximately $0.25$--$0.32$ G. This branch therefore requires a weaker poloidal field than the $3$--$7$-G estimates above. Its period is comparable to the observed Gleissberg activity band near $88$--$90$ yr~\cite{Knudsen2009,SolarLong2019}, although that variability does not establish a magnetic polarity reversal on this timescale. We return below to the possible connection between this longer dynamo wave and a near-203-year modulation.

For the primary cycle, fixing the reversal at 11 yr turns the field and flow measurements into constraints on shear and density. Eqs.~(\ref{solar_orbit_relations}) and~(\ref{solar_circular_local_scaling}) give
\begin{equation}
\boxed{\begin{gathered}
\frac{T_{\mathrm B}}2=11\,\mathrm{yr}
\quad\Longrightarrow\quad
\omega_{\mathrm c}=\frac{\pi}{T_{\mathrm B}}
\simeq4.5\times10^{-9}\,\mathrm{s^{-1}};
\\[7pt]
\frac{\Omega}{2\pi}=443\,\mathrm{nHz},\qquad
T_{\mathrm{orb}}=\frac{2\pi}{\Omega}\simeq26\,\mathrm{days}
\\
\Longrightarrow\quad
\frac{B_{\phi,\max}}{B_{\mathrm{lat},\max}}
=\frac{q\Omega}{\omega_{\mathrm c}}
=\frac{4q}{T_{\mathrm{orb}}}\frac{T_{\mathrm B}}2
\simeq615q;
\\[7pt]
B_{\mathrm{lat},\max}=|V|\sqrt{4\pi\rho}\sqrt{\frac{2-q}{q}}
\\
\Longrightarrow\quad
\rho=\frac{B_{\mathrm{lat},\max}^{\,2}}{4\pi V^2}\frac{q}{2-q}.
\end{gathered}}
\label{solar_shallow_candidate}
\end{equation}
Table~\ref{tab:nssl_constraints} combines these relations with stability and the Model S density profile. The inferred depth must also supply the required helioseismic shear; since the diagnostics sample different locations, these are regional constraints.

Allowing factors-of-a-few agreement extends this comparison to the fiducial 1-Mm depth, where $\rho=2.1\times10^{-6}\,\mathrm{g\,cm^{-3}}$. With $V=10\,\mathrm{m\,s^{-1}}$, the allowed $q\simeq0.5$--$1.1$ gives $5$--$9$ G poloidally and $2.7$--$3.2$ kG toroidally. Fig.~\ref{fig:cycles} shows the dynamo wave for $q=0.506$: all three components reverse every 11 yr with steady circulation, a field ratio near $310$, and poloidal and toroidal amplitudes near $9$ G and $2.8$ kG. These are on the observed polar-field scale and within factors of a few of Antia et al.'s latitudinal and toroidal estimates; the associated modulation is discussed below. At this shear and speed, depths of 0.9--1.4 Mm have Model S densities of $1.8\times10^{-6}$--$4.3\times10^{-6}\,\mathrm{g\,cm^{-3}}$, giving $8$--$13$ G poloidally and $2.5$--$4$ kG toroidally~\cite{ModelS}.

Greater depth makes the comparison harder because the amplitudes scale as $|V|\sqrt{\rho}$. At 14 Mm, $\rho\simeq1.4\times10^{-3}\,\mathrm{g\,cm^{-3}}$ and the inferred $16\,\mathrm{m\,s^{-1}}$ flow~\cite{ModelS,BasuAntiaFlow2010} give $B_{\phi,\max}\simeq132\sqrt{q(2-q)}$ kG: approximately $114$--$132$ kG over the allowed stable shear range, with $370$ G poloidally and $115$ kG toroidally at $q=0.506$. Matching the inferred $1$--$2$ kG instead requires either $q\simeq3\times10^{-5}$--$10^{-4}$, where the wave is unstable, or similarly small $2-q$, where it is stable but the field ratio approaches $1230$, above the inferred $140$--$670$. Thus the period, field, flow, and stability constraints jointly favor moderate shear in the shallow NSSL.

Can the oscillating-circulation family also represent the $11$ yr cycle? Applying it to the same NSSL parameters gives $T_{\rm osc}\simeq1.25$ yr, $V_{\max}\simeq880\,\mathrm{m\,s^{-1}}$, and a toroidal amplitude near $0.34$ kG from Eqs.~(\ref{elliptic_solution}) and~(\ref{elliptic_fields}). Its meridional flow reverses and its shear remains near $S_x/\Omega=-2$, unlike the steady poleward flow and moderate shear of the successful dynamo-wave example. This orbit is also unstable at the fiducial field gradient. Appendix~\ref{app:oscillating} finds bounded nontrivial disturbances only for $0.161<b_{\mathrm p}/\Omega<0.493$, corresponding to periods of $0.084$--$0.158$ yr at the solar rotation rate; in the weak-field limit, a growing disturbance increases by a factor approaching $5.1$ per recurrence. The oscillating-circulation family therefore cannot reproduce the primary solar cycle while satisfying the field, flow, shear, and stability constraints.

\subsection{Long-period modulation of the dynamo wave}\label{sec:modulation}
The dynamo waves considered above can also support slower variations of their circulation, shear, and magnetic field. We now examine how these modulation periods depend on shear and amplitude, comparing them with the Gleissberg and Suess--de Vries activity bands near $88$--$90$ and $208$--$220$ yr~\cite{Knudsen2009,Kim2020,SolarLong2019}.

Within the stable regime, linear perturbations around the dynamo wave oscillate at frequencies given by the dispersion relation in Eq.~(\ref{fixed_magnetized_dispersion}). The slow and fast frequencies satisfy $\omega_\pm^2=(A_2\pm\sqrt{A_2^2-4A_0})/2$, with the minus and plus signs, respectively. Their periods are $T_\pm=2\pi/\omega_\pm$. In the weak-gradient limit (small $b_{\mathrm p}/\Omega$), the dispersion relation gives
\begin{equation}
\boxed{T_-=\frac{2\pi}{\omega_-}\simeq T_{\mathrm B}\sqrt{\frac{2q}{2q-1}}.}
\label{solar_modulation_periods}
\end{equation}
To obtain the finite-amplitude modulation periods, we numerically solve the circulation--shear--tension equations.

Fig.~\ref{fig:phase}(c,d) shows the $q$ dependence of these slow modulation periods for reference reversal periods of 11 and 88 yr, respectively, at $\Omega/(2\pi)=443$ nHz and steady speed $V_0=10\,\mathrm{m\,s^{-1}}$. At each $q$, the reversal period fixes $\omega_{\mathrm c}$, and the steady wave determines $b_{\mathrm p},C_1,C_2$. These quantities remain fixed along the corresponding finite-amplitude trajectories. The color map shows different velocity semi-amplitudes $V_{\rm amp}=(V_{\max}-V_{\min})/2=0$, $1$, $2$, $3$, and $3.5\,\mathrm{m\,s^{-1}}$, with zero denoting the linear limit.

For the 11-year wave, panel~(c) shows the slow period decreasing from centuries near $q=1/2$ to approximately $30$ yr at $q=1.1$ and $27$ yr at $q=1.4$. The fast period remains approximately $21$--$34$ days, comparable to rotation, so the long modulations belong to the slow mode. The 11-year dynamo wave admits a continuous range of modulation periods that includes the observed activity bands. However, the present model does not favor the particular shears that yield 88 or 208 yr, so these periods are possible outcomes rather than preferentially selected timescales.

The solar phase portrait in panel~(b) illustrates these modulations around a steady $11$ yr dynamo wave, marked by the gold star. Here $q=0.50565$ is chosen to give a small-amplitude period near 208 yr; the crimson trajectory illustrates this choice rather than a preferred timescale. Nested modulation curves surround the steady wave, with no secondary island chain resolved. The black cross marks an unstable periodic orbit: circulation, shear, and tension repeat in time, but small disturbances can grow. Trajectories approaching the dashed separatrix spend progressively longer near this orbit, lengthening their modulation period. At this $(C_1,C_2)$, shear and tension keep growing beyond the separatrix.

Panel~(d) shows the case of a longer period 88-year reversal. Its slow modulation decreases to approximately $220$ yr at $q=1.4$ and $207$--$203$ yr over $q=1.8$--$1.999$, with almost no dependence on the displayed amplitudes. This weak shear dependence follows from Eq.~(\ref{solar_modulation_periods}), whose strong-shear estimate is $T_-\simeq2T_{\mathrm B}/\sqrt{3}\simeq203$ yr for $T_{\mathrm B}=176$ yr. Thus the connection between an 88-year reversal and a near-203-year modulation does not require precise shear tuning, although the field and flow constraints favor $q$ closer to 2 (Appendix~\ref{app:long_reversal}). Such a wave could occur separately from the moderate-shear 11-year wave and contribute to long-period activity variability, but an 88-year polarity reversal and the coexistence of these waves remain unestablished.

The non-solar portraits illustrate the wider range of recurring motions. Panel~(e) shows hydrodynamic circulation coupled to zonal flow, with the separatrix dividing circulation that preserves or reverses its direction; a purely toroidal field can be passively advected. Panel~(f) shows magnetic islands separated by separatrices. Its gold diamond is a steady dynamo wave, whereas the cyan centers and black saddles represent stable and unstable periodic motions of circulation, shear, and tension. These are periodic points of the Poincar\'e map, rather than equilibria of the circulation--shear--tension equations.

The magnetic modulation amplitudes provide a separate observational test. We fit the 11-year running mean of the annual SILSO sunspot number over 1700--2025 with
\begin{equation}
\begin{aligned}
N_{\mathrm{SSN}}(t)={}&N_0+N_1(t-t_0)
+a_{\mathrm P}\cos\frac{2\pi t}{\mathcal P}
+c_{\mathrm P}\sin\frac{2\pi t}{\mathcal P},\\
f_{\mathrm P}={}&
\frac{\sqrt{a_{\mathrm P}^2+c_{\mathrm P}^2}}{N_0},
\end{aligned}
\label{solar_band_fit}
\end{equation}
where $N_0$ is the baseline at $t_0=1863$ yr and $N_1$ its trend. The fractional semi-amplitudes $f_{208}\simeq14\%$ and $f_{88}\simeq24\%$ depend on record length, especially for the longer band~\cite{SILSO}.

For comparison, small disturbances on each steady wave's fixed-$C_1,C_2$ surface give a slow toroidal-envelope semi-amplitude
\begin{equation}
f_{B,-}\simeq
\frac{\omega_{\mathrm c}}{\Omega}
\frac{V_{\rm amp}}{|V_0|}
\frac{1}{\sqrt{(2-q)(2q-1)}},
\label{solar_modulation_amplitude}
\end{equation}
in the weak-gradient limit with separated fast and slow frequencies. Here $V_0$ is the reference speed and $V_{\rm amp}=(V_{\max}-V_{\min})/2$ includes both modes, whose amplitudes are linked by the conserved quantities.

With $V_0=4$--$20\,\mathrm{m\,s^{-1}}$ and $V_{\rm amp}\leq3.5\,\mathrm{m\,s^{-1}}$, the tested 11-year-wave continuations over $0.5<q<1.1$ reach approximately $0.5\%$ modulation at 88--90 yr and $0.9\%$ at 208--220 yr. For the 88-year wave over $1.4<q<2$, near-203-year modulation reaches approximately $0.5\%$ while retaining kilogauss toroidal fields at $\rho\leq2.6\times10^{-5}\,\mathrm{g\,cm^{-3}}$. Moving closer to $q=2$ increases modulation but weakens the background fields and eventually shifts the periods; near $q=1/2$, the bounded region contracts and the periods lengthen. These families therefore reproduce the long timescales more readily than the activity amplitudes, although activity need not respond linearly to the mean field. This comparison remains indirect because magnetic-field and meridional-flow records do not yet span either long cycle~\cite{SolarFlow2011,Imada2020,HathawayNSSL2022}.

\subsection{A tachocline source}

Having tested the NSSL wave, we apply the same field, flow, and stability constraints to a tachocline source, with magnetic flux subsequently transported to the surface. An 11-year reversal is not directly measured there. The low-latitude radial shear has the opposite sign to this reversal family; positive shear occurs at higher latitudes, with its magnitude dependent on the inferred transition profile~\cite{AntiaBasuTach2011,BasuTach2026,KorzennikEffDarwich2026}.

The main constraint comes from the much higher density. Taking $\rho\simeq0.19\,\mathrm{g\,cm^{-3}}$, $\Omega/(2\pi)=410$ nHz, and an equatorward speed of $3$--$4\,\mathrm{m\,s^{-1}}$~\cite{ModelS,Gizon}, Eq.~(\ref{solar_circular_local_scaling}) gives, for an 11-year reversal,
\begin{equation}
\begin{aligned}
B_{r,\max}=B_{\mathrm{lat},\max}
&\simeq(0.46\text{--}0.62)
\sqrt{\frac{2-q}{q}}\,\mathrm{kG},\\
B_{\phi,\max}
&\simeq(264\text{--}352)
\sqrt{q(2-q)}\,\mathrm{kG}.
\end{aligned}
\label{solar_local_tach_candidate}
\end{equation}
Thus $q=0.5$--$1.8$ requires toroidal amplitudes of approximately $160$--$350$ kG, far above the $1.4$--$6.5$-kG peaks reconstructed from surface magnetograms and helioseismic rotation~\cite{SolarInterior2026}. Even $q=1.99$ gives $37$--$50$ kG. Reaching the reconstructed range requires either $q\lesssim3\times10^{-4}$, where the wave is unstable, or $2-q\lesssim3\times10^{-4}$. The stable part of the latter range has small-amplitude modulation periods of only about $20$--$25$ yr. Adjusting shear therefore trades weaker fields against the century-scale modulation. These field reconstructions are model dependent: the conditional seismic upper limit of $300$ kG for a layer of half-thickness $0.02R_\odot$ still permits strong-field solutions~\cite{SolarFieldLimit2000}. Nevertheless, the shallow NSSL satisfies the field, flow, and period constraints more readily.

This local analysis does not exclude a global cycle connecting the tachocline and surface. The constant-density approximation applies only over distances small compared with the density scale height and other background variation scales, so its velocity gradient cannot be extrapolated across the convection zone. A global calculation requires stratification, spatially varying shear and circulation, and magnetic transport. It must also extend the meridional force balance: here $\dot\omega_{\mathrm c}=-\Omega S_z$ requires $\langle S_z\rangle=0$ for bounded circulation, whereas the Sun generally has nonzero mean vertical shear~\cite{SolarMeridionalBalance2011}. Helioseismic evidence of cyclic rotational gradients near the tachocline further motivates testing such a global description~\cite{MandalKosovichev2026}.

\section{Conclusion}
Large-scale magnetic cycles can emerge from the non-linear interplay of circulation, shear, and magnetic tension in rotating flows. We obtain exact solutions to incompressible MHD in background rotational shear by taking the velocity and magnetic fields to vary linearly with position, preserving their non-linear interactions. The resulting feedback follows the cycle shown in Fig.~\ref{fig:feedback}:
\begin{enumerate}
\item \textit{Meridional circulation rotates the poloidal field.}
Induction turns the radial and vertical field components into one another at $2\omega_{\mathrm c}$, changing the field on which shear acts.

\item \textit{Shear stretches the poloidal components into toroidal field.}
Radial shear stretches the radial field azimuthally, while vertical shear does the same to the vertical field.

\item \textit{The resulting toroidal field acts back on the shear through magnetic tension.}
Together with the poloidal field, it exerts azimuthal tension that changes both radial and vertical shear.

\item \textit{The changed vertical shear alters the meridional circulation.}
Vertical shear produces different radial Coriolis accelerations at different heights. Together with pressure-driven vertical motion, this changes the circulation rate, completing the feedback loop.
\end{enumerate}
Circulation also changes shear directly: advection converts radial and vertical shear into one another, while Coriolis deflection of the radial flow generates vertical shear because the radial velocity varies with height.

Two conserved quantities reduce the dynamics to three coupled equations. The first fixes the difference between squared velocity and magnetic gradients, corresponding to kinetic minus magnetic energy in a symmetric patch; the second couples absolute vorticity to the poloidal field and current-helicity gradients. We classify the steady-state families, construct exactly integrable non-linear orbits such as the steady- and oscillating-circulation magnetic cycles, and determine their stability. The key solution, the steady-circulation reversal orbit or dynamo wave, maintains constant velocity and tension while all magnetic components undergo a periodic reversal.

The toroidal-to-poloidal ratio sets the pace of this reversal through $T_{\mathrm B}/2=(T_{\mathrm{orb}}/4q)(B_{\phi,\max}/B_{\mathrm{lat},\max})$. Toroidal dominance leads to a cycle period many times the rotation period. Shear acting on a weak poloidal field changes only a small fraction of the stronger toroidal field per rotation. The weak poloidal field also exerts less magnetic torque, whose balance against advection and Coriolis forces requires slower circulation at fixed shear and rotation. Circulation therefore turns the poloidal field gradually, reversing toroidal induction over many rotations. Efficient regeneration of poloidal flux would shorten the cycle period. Since shear naturally yields a large toroidal-to-poloidal ratio, a long period cycle such as the 11 yr solar cycle favors shear as its primary driver. Convection/buoyancy is probably a secondary actor, mainly responsible for transporting and replenishing magnetic flux.

Our solar application tests the physical plausibility of this mechanism rather than seeking a precise fit to the data. The period, field, flow, and stability constraints jointly favor the 11-year dynamo wave to operate at moderate shear, approximately $0.5\lesssim q\lesssim1.1$, in the shallow near-surface shear layer (NSSL). At a depth near 1 Mm, the allowed shear range $q\simeq0.5$--$1.1$ and observed poleward speeds of $8$--$12\,\mathrm{m\,s^{-1}}$ give stable 11-year reversals with poloidal amplitudes of approximately $4$--$11$ G and toroidal amplitudes of $2.2$--$3.8$ kG, in reasonable agreement with regional observations.

This 11-year wave also supports slower modulations of circulation, shear, and magnetic tension. Their periods decrease from centuries near $q=1/2$ to approximately $30$ yr at $q=1.1$, encompassing the observed Gleissberg and Suess--de Vries bands. These longer timescales are therefore possible around the primary reversal, although the model does not preferentially select 88 or 208 yr.

Stronger shear offers a distinct connection between the long periods. An 88-year dynamo wave supports slow modulation near $203$--$207$ yr over $q=1.8$--$1.999$, without precise shear tuning. The 88-year reversal and near-203-year modulation thus fit together naturally as a primary magnetic cycle and its slower variation. Matching $1$--$3$-kG toroidal fields and $8$--$12\,\mathrm{m\,s^{-1}}$ poleward flow at the 1-Mm density requires $q\simeq1.989$--$1.9995$ and $0.1$--$0.3$-G poloidal fields, weaker than those of the 11-year wave. The magnetic energy repeats every 88 yr and could contribute to long-period activity variability, although an 88-year polarity reversal and coexistence with the 11-year wave remain unestablished. For the trajectories studied on each steady wave's fixed-$C_1,C_2$ surface, the slow toroidal-envelope fractional semi-amplitudes remain below approximately $1\%$ in the relevant bands, compared with $14$--$24\%$ in sunspot number; activity need not respond linearly to the mean field though.

At the much higher tachocline density, the inferred return flow requires substantially stronger fields for the same 11-year reversal. Over $q=0.5$--$1.8$, the predicted toroidal amplitudes are approximately $160$--$350$ kG, well above recent large-scale-field reconstructions. Moving sufficiently close to $q=2$ reduces the fields but shortens the small-amplitude modulation to decades. Strong-field solutions remain possible under the broader seismic bound. Our analysis suggests that the shallow NSSL satisfies the joint constraints more readily than the tachocline. The possibility of a global solar cycle spanning the convection zone is, however, left open.

Our affine solution describes coherent large-scale fields but omits interacting small-scale fluctuations, their turbulent cascade, and their back-reaction. Follow-up work will introduce coupled spatial modes, including sinusoidal variations, and derive the phase-synchronization conditions necessary for long period cycles. Stratification, convection/buoyancy, and mechanisms such as the Tayler--Spruit dynamo would have to be included to address poloidal-field maintenance against diffusion. Ultimately, settling the debate regarding whether the solar dynamo is local or global requires self-consistent simulations of the solar convection zone, including the NSSL, with stratification, realistic rotational shear, and meridional flow. These simulations must test whether dynamo cycles and steady meridional flows persist, and establish the relative roles of shear, convection, and buoyancy. By identifying the feedback that drives magnetic reversals in shear flows, this work provides a physical foundation for these tests and for understanding large-scale cyclic dynamos throughout the Universe.

\section*{Acknowledgments}
The author thanks Bindesh Tripathy, Amitava Bhattacharjee, James Stone, and George Wong for insightful discussions and valuable suggestions. OpenAI Codex, running GPT-5.6 Sol and GPT-6 Astra, assisted with algebra, literature synthesis, numerical code, figures, and editing; all scientific decisions and verification remain the author's responsibility. Including delegated research tasks, the recorded Codex usage was approximately $7\times10^8$ input and output tokens, including cached input, over about ten days.

\let\section\appendixsection
\let\subsection\appendixsubsection
\allowdisplaybreaks[1]
\appendix
\setcounter{secnumdepth}{2}
\section{Derivation of the coefficient equations}\label{app:reduction}
The solutions in the main text follow from the coefficient equations and conserved quantities derived here; these also provide the dynamics for the steady states and time-dependent orbits below. For the fields in Eq.~(\ref{fields}), both the fields and pressure are independent of $y$. Define $D_t=\partial_t+U_x\partial_x+U_z\partial_z$, the derivative following the fluid. Eqs.~(\ref{momentum})--(\ref{MHD}) then read, component by component,
\begin{align}
D_tU_x-2\Omega U_y
&=-\partial_xP+B_x\partial_xB_x+B_z\partial_zB_x\nonumber\\
&\quad+2q\Omega^2x,\\
D_tU_y+2\Omega U_x
&=B_x\partial_xB_y+B_z\partial_zB_y,\\
D_tU_z&=-\partial_zP+B_x\partial_xB_z+B_z\partial_zB_z,\\
D_tB_x&=B_x\partial_xU_x+B_z\partial_zU_x,\\
D_tB_y&=B_x\partial_xU_y+B_z\partial_zU_y,\\
D_tB_z&=B_x\partial_xU_z+B_z\partial_zU_z.
\label{app:components}
\end{align}
Here $\nabla^2=\partial_x^2+\partial_z^2$. Eq.~(\ref{fields}) satisfies both divergence constraints and has zero vector Laplacians, so the viscous and resistive terms vanish exactly.

For radial induction, direct differentiation gives
$\boldsymbol U\cdot\nabla B_x=\omega_{\mathrm c}b_{xz}x-
\omega_{\mathrm c}b_{xx}z$ and
$\boldsymbol B\cdot\nabla U_x=-\omega_{\mathrm c}b_{xz}x+
\omega_{\mathrm c}b_{xx}z$.  Substitution in $D_tB_x-\boldsymbol B\cdot\nabla U_x=0$ leaves
$(\dot b_{xx}+2\omega_{\mathrm c}b_{xz})x+
(\dot b_{xz}-2\omega_{\mathrm c}b_{xx})z$; setting both independent
coefficients to zero gives the poloidal pair in Eq.~(\ref{seven}).
The vertical induction equation gives the same two conditions independently.

For azimuthal induction, the coefficients of $x,z$ in $D_tB_y$ are
$\dot b_{yx}+\omega_{\mathrm c}b_{yz}$ and
$\dot b_{yz}-\omega_{\mathrm c}b_{yx}$; those in $\boldsymbol B\cdot\nabla U_y$ are
$S_xb_{xx}+S_zb_{xz}$ and $S_xb_{xz}-S_zb_{xx}$.  Matching each pair gives
the last two induction equations in Eq.~(\ref{seven}).  Likewise,
$D_tU_y+2\Omega U_x$ has coefficients
$\dot S_x+\omega_{\mathrm c}S_z$ and
$\dot S_z-\omega_{\mathrm c}(S_x+2\Omega)$.  Equating them to the tension
coefficients $\mathcal T_x,\mathcal T_z$ gives the two shear equations.

The magnetic and shear equations leave the circulation to be determined by meridional momentum. The radial and vertical magnetic tensions reduce to $b_{\mathrm p}^2x$ and
$b_{\mathrm p}^2z$.  The meridional advection terms are
$\boldsymbol U\cdot\nabla U_x=-\omega_{\mathrm c}^2x$ and
$\boldsymbol U\cdot\nabla U_z=-\omega_{\mathrm c}^2z$, while the time
derivatives are $-\dot\omega_{\mathrm c}z$ and
$\dot\omega_{\mathrm c}x$.  Substitution in radial and vertical momentum,
including the Coriolis force, gives
\begin{align}
\partial_xP={}&(b_{\mathrm p}^2+\omega_{\mathrm c}^2+2\Omega S_x+2q\Omega^2)x\nonumber\\
&+(\dot\omega_{\mathrm c}+2\Omega S_z)z,\\
\partial_zP={}&-\dot\omega_{\mathrm c}x+(b_{\mathrm p}^2+\omega_{\mathrm c}^2)z.
\end{align}
The radial and vertical pressure gradients must satisfy $\partial_z(\partial_xP)=\partial_x(\partial_zP)$, since the order of spatial differentiation does not change the result. Differentiating the two expressions above therefore gives $\dot\omega_{\mathrm c}+2\Omega S_z=-\dot\omega_{\mathrm c}$, or $\dot\omega_{\mathrm c}=-\Omega S_z$, completing
Eq.~(\ref{seven}).  Integrating gives Eq.~(\ref{pressure}); adding $\partial_x^2P$ and $\partial_z^2P$ gives
\begin{equation}
\nabla^2P=2q\Omega^2+2\Omega S_x+2\omega_{\mathrm c}^2+2b_{\mathrm p}^2.
\label{app:pressure_laplacian}
\end{equation}
For any bounded periodic affine motion, integrating
$\dot\omega_{\mathrm c}=-\Omega S_z$ over one period gives
$\langle S_z\rangle=0$, where the brackets denote an average over that period.

With all seven coefficients determined, the azimuthal magnetic force can be followed directly. Applying the product rule to its gradients and using the four induction equations gives
\begin{equation}
\begin{aligned}
\dot{\mathcal T}_x
&=\dot b_{yx}b_{xx}+b_{yx}\dot b_{xx}
+\dot b_{yz}b_{xz}+b_{yz}\dot b_{xz}\\
&=b_{\mathrm p}^2S_x-\omega_{\mathrm c}\mathcal T_z,\\
\dot{\mathcal T}_z
&=\dot b_{yx}b_{xz}+b_{yx}\dot b_{xz}
-\dot b_{yz}b_{xx}-b_{yz}\dot b_{xx}\\
&=b_{\mathrm p}^2S_z+\omega_{\mathrm c}\mathcal T_x.
\end{aligned}\label{app:tension_derivatives}
\end{equation}
These identities yield Eq.~(\ref{five}). To derive its constants, multiply the two shear equations by $S_x,S_z$, and the two tension equations by $\mathcal T_x,\mathcal T_z$. Addition within each pair gives
\begin{align}
\tfrac12\frac{\dd}{\dd t}(S_x^2+S_z^2)
&=S_x\mathcal T_x+S_z\mathcal T_z+2\Omega\omega_{\mathrm c}S_z,\\
\frac{1}{2b_{\mathrm p}^2}\frac{\dd}{\dd t}(\mathcal T_x^2+\mathcal T_z^2)
&=S_x\mathcal T_x+S_z\mathcal T_z.
\end{align}
Subtracting and using $\dd(\omega_{\mathrm c}^2)/\dd t=-2\Omega\omega_{\mathrm c}S_z$ yields $C_1$. For $C_2$, the product rule gives
\begin{equation}
\begin{aligned}
\frac{\dd}{\dd t}(S_z\mathcal T_x-S_x\mathcal T_z)
&=2\Omega\omega_{\mathrm c}\mathcal T_x\\
&=\frac{\dd}{\dd t}(2\Omega\mathcal T_z+2b_{\mathrm p}^2\omega_{\mathrm c}).
\end{aligned}
\end{equation}
Integration gives the second constant in Eq.~(\ref{constants}). Solving the two constant relations for the shears gives the expressions used in Eqs.~(\ref{root}) and~(\ref{three}). This reduction identifies the independent evolving quantities; the next step is to find their steady balances and response to disturbances.

\section{Fixed points and linear stability}\label{app:stability}
Steady circulation, shear, and magnetic tension require the five right-hand sides of Eq.~(\ref{five}) to vanish. The circulation and radial-shear equations give
$S_z=\mathcal T_x=0$.  The remaining equations require
$\mathcal T_z=-\omega_{\mathrm c}(S_x+2\Omega)$ and
$b_{\mathrm p}^2S_x=\omega_{\mathrm c}\mathcal T_z$, hence
$(b_{\mathrm p}^2+\omega_{\mathrm c}^2)S_x=
-2\Omega\omega_{\mathrm c}^2$.
For $\omega_{\mathrm c}\ne0$, these give Eq.~(\ref{fixed_curve}). If $\omega_{\mathrm c}=0$, then $\mathcal T_z=0$, with $S_x=0$ for $b_{\mathrm p}>0$ and arbitrary $S_x$ for $b_{\mathrm p}=0$. This proves that the four classes in the main text exhaust the reduced fixed points.

To test these balances, the linear-response matrix, or Jacobian, records how each rate changes when one variable is perturbed. At fixed $b_{\mathrm p}$, differentiating Eq.~(\ref{five}) with respect to
$(\omega_{\mathrm c},S_x,S_z,\mathcal T_x,\mathcal T_z)$ gives
\begin{equation}
J_{\mathcal T}=\begin{pmatrix}
0&0&-\Omega&0&0\\
-S_z&0&-\omega_{\mathrm c}&1&0\\
S_x+2\Omega&\omega_{\mathrm c}&0&0&1\\
-\mathcal T_z&b_{\mathrm p}^2&0&0&-\omega_{\mathrm c}\\
\mathcal T_x&0&b_{\mathrm p}^2&\omega_{\mathrm c}&0
\end{pmatrix}.
\label{app:jacobian5}
\end{equation}
For a disturbance proportional to $e^{-i\omega t}$, the frequencies satisfy $\det(-i\omega I-J_{\mathcal T*})=0$, where $I$ is the identity matrix and the star means evaluation on the steady solution. For the dynamo wave (the steady-circulation reversal orbit), substitute $S_x=-q\Omega$,
$S_z=\mathcal T_x=0$, $\mathcal T_z=-(2-q)\Omega\omega_{\mathrm c}$,
and $b_{\mathrm p}^2=(2-q)\omega_{\mathrm c}^2/q$. Direct expansion gives
\begin{equation}
\det(-i\omega I-J_{\mathcal T*})
=-i\omega(\omega^4-A_2\omega^2+A_0),
\end{equation}
where $A_2,A_0$ are given in Eq.~(\ref{fixed_magnetized_dispersion}). The zero root moves along the fixed-point curve. To classify the other roots, write $\xi=b_{\mathrm p}^2/\Omega^2$. The two values of $\omega^2$ are real when $A_2^2-4A_0\geq0$. The nonzero boundaries at which a value passes through zero or the two values merge are, respectively,
\begin{equation}
\xi_0=\frac{(1-2q)(2-q)^2}{2},\qquad
\xi_{\mathrm m}=\frac{2-q}{4}\left(\sqrt{\frac{2}{q}}-1\right).
\label{app:stability_boundaries}
\end{equation}
The first boundary is positive only for $q<1/2$. Below $\xi_0$, the product of the two values of $\omega^2$ is negative, so one frequency pair is real and the other imaginary. Between $\xi_0$ and $\xi_{\mathrm m}$, both values are positive for $q>(3-\sqrt5)/4$ and negative for smaller $q$. Above $\xi_{\mathrm m}$, the two values of $\omega^2$ are complex conjugates. These sign tests give every case in Eq.~(\ref{stability}).

The same roots also determine the modulation timescale around a stable reversal. The smaller oscillatory root and its modulation period $T_-$ are
\begin{equation}
\omega_-^2=\frac12\left[A_2-\sqrt{A_2^2-4A_0}\right],
\qquad
T_-=\frac{T_{\rm orb}}{\omega_-/\Omega}.
\label{app:slow_root}
\end{equation}
For the fiducial $q=0.506$ and $b_{\mathrm p}/\Omega\simeq0.00279$, the base orbit reverses in 11 yr and Eq.~(\ref{app:slow_root}) gives a small-amplitude modulation near 202 yr. Fig.~\ref{fig:phase}(b) instead uses $q=0.50565$, with the conserved quantities fixed to those of the steady 11-year wave, giving a small-amplitude modulation near 208 yr. The section $S_z=0$, $\dot S_z>0$ samples on the fast timescale and resolves this slow modulation.

The other families require attention to the individual magnetic coefficients, because constant shear and tension can coexist with a changing magnetic field. Setting every derivative in Eq.~(\ref{seven}) to zero requires $S_z=0$ and $\omega_{\mathrm c}b_{\mathrm p}=0$. If $b_{\mathrm p}>0$, induction gives $\omega_{\mathrm c}=S_x=0$, and Eq.~(\ref{tensions}) then requires $b_{yx}=b_{yz}=0$. If $b_{\mathrm p}=0$, either $\omega_{\mathrm c}=0$ with arbitrary $S_x,b_{yx},b_{yz}$, or $\omega_{\mathrm c}\ne0$ with $S_x=-2\Omega$ and zero magnetic coefficients. Thus the steady-circulation reversal and the toroidal-rotation family with a finite toroidal field have constant circulation, shear, and tension, while their magnetic fields vary periodically.

The Jacobian $J$ is obtained by differentiating the seven right-hand sides
with respect to $(\omega_{\mathrm c},S_x,S_z,b_{xx},b_{xz},b_{yx},b_{yz})$.
Grouping velocity and magnetic variables gives the four blocks
\begin{subequations}\label{app:jacobian7}
\begin{align}
J&=\begin{pmatrix}J_{\mathrm{UU}}&J_{\mathrm{UB}}\\J_{\mathrm{BU}}&J_{\mathrm{BB}}\end{pmatrix},\\\displaybreak[1]
J_{\mathrm{UU}}&=\begin{pmatrix}
0&0&-\Omega\\-S_z&0&-\omega_{\mathrm c}\\S_x+2\Omega&\omega_{\mathrm c}&0
\end{pmatrix},\\\displaybreak[1]
J_{\mathrm{UB}}&=\begin{pmatrix}
0&0&0&0\\b_{yx}&b_{yz}&b_{xx}&b_{xz}\\-b_{yz}&b_{yx}&b_{xz}&-b_{xx}
\end{pmatrix},\\\displaybreak[1]
J_{\mathrm{BU}}&=\begin{pmatrix}
-2b_{xz}&0&0\\2b_{xx}&0&0\\-b_{yz}&b_{xx}&b_{xz}\\b_{yx}&b_{xz}&-b_{xx}
\end{pmatrix},\\\displaybreak[1]
J_{\mathrm{BB}}&=\begin{pmatrix}
0&-2\omega_{\mathrm c}&0&0\\2\omega_{\mathrm c}&0&0&0\\S_x&S_z&0&-\omega_{\mathrm c}\\-S_z&S_x&\omega_{\mathrm c}&0
\end{pmatrix}.
\end{align}
\end{subequations}
The labels $\mathrm U,\mathrm B$ identify the velocity and magnetic groups. Let $\boldsymbol\delta$ collect the seven coefficient perturbations. For their $e^{-i\omega t}$ dependence, linearization gives $(-i\omega I-J_*)\boldsymbol\delta=0$, with the same identity-matrix and equilibrium notation as above. A nonzero mode requires a zero determinant. The subscripts $\mathrm{pol},\mathrm{shear},\mathrm{circ}$ below denote the stationary poloidal, stationary shear--toroidal, and zero-toroidal-amplitude member of the toroidal-rotation family. Substituting their equilibrium values into these blocks gives
\begin{align}
\det(-i\omega I-J_{\mathrm{pol}})
&=i\omega^3(\omega^2+b_{\mathrm p}^2)
(\omega^2+b_{\mathrm p}^2-2\Omega^2)=0,\\
\det(-i\omega I-J_{\mathrm{shear}})
&=i\omega^5[\omega^2-\Omega(S_x+2\Omega)]=0,\\
\det(-i\omega I-J_{\mathrm{circ}})
&=i\omega(\omega^2-\omega_{\mathrm c}^2)^2
(\omega^2-4\omega_{\mathrm c}^2)=0.
\label{app:determinants}
\end{align}
Dividing by $i$ gives Eqs.~(\ref{fixed_poloidal}), (\ref{fixed_shear}), and~(\ref{fixed_center}), respectively. The poloidal family has $\omega=\pm i b_{\mathrm p}$ and is unstable.

The repeated roots in these dispersion relations require one further check: whether independent disturbances share the frequency or drive one another. The latter case produces growth proportional to time. An eigenvector specifies a disturbance pattern that changes only by an overall factor. A size-two Jordan block is a two-by-two part of the response matrix describing two coupled directions with only one independent eigenvector. For eigenvalue $\lambda=-i\omega$, it has the form
\begin{equation}
\begin{aligned}
J_\lambda&=\lambda I+N,\qquad N\ne0,\qquad N^2=0,\\
e^{J_\lambda t}
&=e^{\lambda t}e^{Nt}
=e^{\lambda t}\left(I+tN+\frac{t^2N^2}{2!}+\cdots\right)\\
&=e^{\lambda t}(I+tN).
\end{aligned}
\label{app:jordan}
\end{equation}
Since $N^2=0$, the series stops at $tN$. Representing an arbitrary initial disturbance requires a second direction: choose $v_1$ with $Nv_1=v_0$ and $Nv_0=0$, where $v_0$ is the eigenvector. This generalized eigenvector feeds the $v_0$ direction, giving
\begin{equation}
e^{J_\lambda t}v_1=e^{\lambda t}(v_1+t v_0).
\label{app:jordan_growth}
\end{equation}
Thus the $v_0$ component grows linearly in time, with $e^{\lambda t}=1$ for the zero root below.

If each root has as many independent eigenvectors as its multiplicity, they form an invertible matrix $\mathcal V$ and give independent modes. For $n$ eigenvalues $\lambda_j$, let $\operatorname{diag}$ denote a matrix with those entries on the diagonal and zeros elsewhere:
\begin{equation}
\begin{aligned}
\mathcal D&=\operatorname{diag}(\lambda_j),\qquad j=1,\ldots,n,\\
J&=\mathcal V\mathcal D\mathcal V^{-1},\\
e^{Jt}&=\mathcal V\,\operatorname{diag}(e^{\lambda_jt})\mathcal V^{-1}.
\end{aligned}
\label{app:diagonalizable}
\end{equation}
A repeated $\lambda$ then repeats the exponential without a $t$ factor: purely imaginary eigenvalues give bounded oscillations, and zero eigenvalues give constants.

The Jordan blocks of the stationary shear--toroidal family are visible directly in Eq.~(\ref{app:jacobian7}). For the member with zero equilibrium toroidal gradients, set $\omega_{\mathrm c}=S_z=b_{xx}=b_{xz}=b_{yx}=b_{yz}=0$. The cross blocks vanish, and the magnetic block in the ordering $(\delta b_{xx},\delta b_{xz},\delta b_{yx},\delta b_{yz})$ becomes
\begin{equation}
J_{\mathrm{BB}}^{\mathrm{shear}}=
\begin{pmatrix}
0&0&0&0\\
0&0&0&0\\
S_x&0&0&0\\
0&S_x&0&0
\end{pmatrix}.
\label{app:shear_jordan}
\end{equation}
Pairing $(\delta b_{xx},\delta b_{yx})$ and $(\delta b_{xz},\delta b_{yz})$ turns this matrix into two identical blocks
\begin{equation}
N_x=\begin{pmatrix}0&0\\S_x&0\end{pmatrix},
\qquad N_x^2=0.
\label{app:shear_block}
\end{equation}
For $S_x\ne0$, each $N_x$ is a zero-eigenvalue Jordan block. The equation $N_xv=0$ forces the first component of $v$ to vanish, leaving only one independent eigenvector, $v_0=(0,1)^{\mathrm T}$. Here the superscript $\mathrm T$ turns the listed entries into a column vector. The second direction may be chosen as $v_1=(1/S_x,0)^{\mathrm T}$, for which $N_xv_1=v_0$. Eq.~(\ref{app:jordan}) gives
\begin{equation}
\begin{aligned}
\delta b_{yx}(t)&=\delta b_{yx}(0)+S_x\delta b_{xx}(0)t,\\
\delta b_{yz}(t)&=\delta b_{yz}(0)+S_x\delta b_{xz}(0)t.
\end{aligned}
\label{app:shear_growth}
\end{equation}
Thus a constant poloidal-gradient perturbation is stretched by the equilibrium shear into a toroidal-gradient perturbation that grows linearly. This is the physical effect of the Jordan block in this family. At $S_x=0$, $N_x=0$ and the four magnetic zero modes are independent constants; together with the neutral shear displacement, they account for the five independent zero modes of the member with zero shear and zero toroidal amplitude.

The velocity block in Eq.~(\ref{app:jacobian7}) contains
\begin{equation}
\frac{\dd}{\dd t}
\begin{pmatrix}\delta\omega_{\mathrm c}\\ \delta S_z\end{pmatrix}
=\begin{pmatrix}0&-\Omega\\ S_x+2\Omega&0\end{pmatrix}
\begin{pmatrix}\delta\omega_{\mathrm c}\\ \delta S_z\end{pmatrix}.
\label{app:shear_velocity_block}
\end{equation}
It oscillates for $S_x>-2\Omega$ and grows exponentially for $S_x<-2\Omega$. At $S_x=-2\Omega$ it has a zero-frequency Jordan block: $\delta S_z$ is constant and $\delta\omega_{\mathrm c}=\delta\omega_{\mathrm c}(0)-\Omega\delta S_z(0)t$. For nonzero equilibrium toroidal gradients, the full matrix instead gives
\begin{equation}
\delta\ddot\omega_{\mathrm c}+\Omega(S_x+2\Omega)\delta\omega_{\mathrm c}
=\Omega(b_{yz}\delta b_{xx}-b_{yx}\delta b_{xz}),
\label{app:shear_forced_circulation}
\end{equation}
where the poloidal disturbances on the right are constant. At $S_x=-2\Omega$, a nonzero right-hand side therefore produces a term proportional to $t^2$ in circulation; integrating the toroidal induction equations then produces terms proportional to $t^3$ in the toroidal coefficients. These higher powers arise only when the corresponding forcing is nonzero.

The toroidal-rotation family provides the contrasting repeated-frequency case. For its member with zero toroidal amplitude, both cross blocks vanish. Its circulation is the nonzero constant $\omega_{\mathrm c}$. The velocity block gives
\begin{equation}
\begin{aligned}
\delta\ddot S_x+\omega_{\mathrm c}^2\delta S_x&=0,\\
\frac{\dd}{\dd t}\left(\delta\omega_{\mathrm c}
-\frac{\Omega}{\omega_{\mathrm c}}\delta S_x\right)&=0.
\end{aligned}
\label{app:rotation_velocity}
\end{equation}
The second line is the constant displacement along the family. The linearized poloidal induction equations are
\begin{equation}
\begin{aligned}
\delta\dot b_{xx}&=-2\omega_{\mathrm c}\delta b_{xz},\\
\delta\dot b_{xz}&=2\omega_{\mathrm c}\delta b_{xx},\\
\delta\ddot b_{xx}+4\omega_{\mathrm c}^2\delta b_{xx}&=0,\\
\delta\ddot b_{xz}+4\omega_{\mathrm c}^2\delta b_{xz}&=0.
\end{aligned}
\label{app:rotation_poloidal}
\end{equation}
This explicitly gives the poloidal frequency $2\omega_{\mathrm c}$. The toroidal equations are
\begin{equation}
\begin{aligned}
\delta\dot b_{yx}&=-2\Omega\delta b_{xx}
-\omega_{\mathrm c}\delta b_{yz},\\
\delta\dot b_{yz}&=-2\Omega\delta b_{xz}
+\omega_{\mathrm c}\delta b_{yx}.
\end{aligned}
\label{app:rotation_toroidal}
\end{equation}
Substituting the poloidal solution at frequency $2\omega_{\mathrm c}$ into these equations gives the particular toroidal response
\begin{equation}
\delta b_{yx}=-\frac{2\Omega}{\omega_{\mathrm c}}\delta b_{xz},
\qquad
\delta b_{yz}=\frac{2\Omega}{\omega_{\mathrm c}}\delta b_{xx}.
\label{app:rotation_response}
\end{equation}
Thus the amplitude of the toroidal-gradient pair is $2\Omega/\omega_{\mathrm c}$ times that of the poloidal pair for $\omega_{\mathrm c}>0$, as quoted in the main text. Equivalently, differentiating Eq.~(\ref{app:rotation_toroidal}) gives
\begin{equation}
\begin{aligned}
\delta\ddot b_{yx}+\omega_{\mathrm c}^2\delta b_{yx}
&=6\Omega\omega_{\mathrm c}\delta b_{xz},\\
\delta\ddot b_{yz}+\omega_{\mathrm c}^2\delta b_{yz}
&=-6\Omega\omega_{\mathrm c}\delta b_{xx}.
\end{aligned}
\label{app:rotation_forcing}
\end{equation}
The forcing at $2\omega_{\mathrm c}$ differs from the natural toroidal frequency $\omega_{\mathrm c}$, so the response is bounded. Independent velocity and toroidal oscillations supply all eigenvectors required by the repeated $\pm\omega_{\mathrm c}$ roots, while the zero mode changes the constant circulation. The repeated roots therefore produce no growth proportional to time when the toroidal-field amplitude is zero.

With nonzero background toroidal amplitude, a small poloidal field produces tension at frequency $\omega_{\mathrm c}$. Eqs.~(\ref{five}) give $\delta\dot{\mathcal T}_x=-\omega_{\mathrm c}\delta\mathcal T_z$ and
\begin{equation}
\delta\ddot S_x+\omega_{\mathrm c}^2\delta S_x
=-2\omega_{\mathrm c}\delta\mathcal T_z.
\end{equation}
The force now oscillates at the natural velocity frequency: this is resonance, and it gives the linear growth stated in the main text.

\section{Analytical nonlinear orbits and their stability}
\label{app:orbits}
The fixed-point analysis describes steady balances between the flow and magnetic tension and their response to small disturbances. To obtain exact motions with changing circulation, shear, and tension, we impose relations that reduce Eqs.~(\ref{five}) to one solvable equation; the remaining quantities then follow by substitution. This gives the analytical orbits discussed in the main text.

\subsection{Oscillating-circulation cycle}\label{app:oscillating}
To retain both magnetic feedback and changing circulation, set $\mathcal T_x=\gamma\omega_{\mathrm c}S_z$, with a constant dimensionless factor $\gamma$ to be determined by the remaining equations. Using
$S_z=-\dot\omega_{\mathrm c}/\Omega$ in the radial-shear and
vertical-tension equations and integrating gives
\begin{equation}
\begin{aligned}
S_x&=S_0+\frac{1-\gamma}{2\Omega}\omega_{\mathrm c}^2,\\
\mathcal T_z&=D_0-\frac{b_{\mathrm p}^2}{\Omega}\omega_{\mathrm c}
-\frac{\gamma}{3\Omega}\omega_{\mathrm c}^3,
\end{aligned}
\end{equation}
where $S_0$ is a constant shear and $D_0$ a constant tension offset. Substitution in the vertical-shear equation gives
\begin{equation}
\ddot\omega_{\mathrm c}+\Omega D_0+
[\Omega(S_0+2\Omega)-b_{\mathrm p}^2]\omega_{\mathrm c}
+\left(\tfrac12-\tfrac{5\gamma}{6}\right)\omega_{\mathrm c}^3=0.
\label{app:general_oscillator}
\end{equation}
Multiplication by $\dot\omega_{\mathrm c}$ makes every term a time derivative, giving a conserved oscillator energy: the sum of a term proportional to $\dot\omega_{\mathrm c}^2$ and terms from the restoring force. Differentiating
$\mathcal T_x=-\gamma\omega_{\mathrm c}\dot\omega_{\mathrm c}/\Omega$ and using this constant to eliminate $\dot\omega_{\mathrm c}^2$ in the remaining radial-tension equation leaves independent constant, linear, quadratic, and quartic powers of $\omega_{\mathrm c}$. Their coefficients require
\begin{equation}
\begin{aligned}
(3\gamma+1)D_0=0,\qquad
4\gamma[\Omega(S_0+2\Omega)-b_{\mathrm p}^2]
=b_{\mathrm p}^2(3-\gamma),\\
\frac32\left(\frac12-\frac{5\gamma}{6}\right)=\frac13.
\end{aligned}
\end{equation}
Thus $\gamma=1/3$, $D_0=0$, and
$S_0=3b_{\mathrm p}^2/\Omega-2\Omega$. The constant term fixes the oscillator energy,
\begin{equation}
\frac12\dot\omega_{\mathrm c}^2+b_{\mathrm p}^2\omega_{\mathrm c}^2
+\frac{\omega_{\mathrm c}^4}{18}
=3\Omega^2b_{\mathrm p}^2-\frac92b_{\mathrm p}^4,
\end{equation}
which yields Eqs.~(\ref{elliptic}) and~(\ref{elliptic_solution}).

Numerically integrating the linearized equations over one cycle gives bounded nontrivial disturbances for $0.161<b_{\mathrm p}/\Omega<0.493$; outside this interval, a disturbance grows exponentially. Neutral disturbances can still accumulate a phase difference proportional to time.

\subsection{Hydrodynamic circulation--shear oscillator}
Removing the poloidal field, $b_{\mathrm p}=\mathcal T_x=\mathcal T_z=0$, isolates the velocity feedback through shear and rotation; any purely toroidal, axisymmetric field is passive because its tension vanishes. The radial-shear equation becomes $\dot S_x=-\omega_{\mathrm c}S_z=\omega_{\mathrm c}\dot\omega_{\mathrm c}/\Omega$.
Integration gives $S_x-\omega_{\mathrm c}^2/(2\Omega)=S_{\mathrm h}$. Substituting
$S_z=-\dot\omega_{\mathrm c}/\Omega$ in the vertical-shear equation gives Eq.~(\ref{hydro_orbits}). Its conserved oscillator energy $E_{\mathrm h}$ is
\begin{equation}
E_{\mathrm h}=\frac12\dot\omega_{\mathrm c}^2+
\frac{\Omega(S_{\mathrm h}+2\Omega)}2\omega_{\mathrm c}^2+
\frac18\omega_{\mathrm c}^4.
\end{equation}
Closed curves around either nonzero center and the sign-changing outer curves are periodic; the energy level through the stationary shear--toroidal saddle forms the boundary between them.

The tension perturbation rotates through the angle $\int\omega_{\mathrm c}\,\dd t$, which is zero over a sign-changing orbit and $+2\pi$ or $-2\pi$ over an orbit with positive or negative circulation, respectively. It therefore returns after each cycle. The velocity period depends on the energy $E_{\mathrm h}$. For two nearby orbits separated by energy $\delta E_{\mathrm h}$, let $\delta\phi_{\mathrm h}$ be their phase difference and $T_{\mathrm h}(E_{\mathrm h})$ the circulation period, with $T'_{\mathrm h}(E_{\mathrm h})=\dd T_{\mathrm h}/\dd E_{\mathrm h}$. They acquire
\begin{equation}
\delta\phi_{\mathrm h}\simeq-\frac{2\pi tT'_{\mathrm h}(E_{\mathrm h})}{T_{\mathrm h}(E_{\mathrm h})^2}\,\delta E_{\mathrm h},
\end{equation}
which grows linearly when $T'_{\mathrm h}(E_{\mathrm h})\ne0$. Thus nearby periodic curves drift out of phase without exponential growth; the boundary orbit is unstable.

\subsection{Families obtained by prescribing the shear}
Both oscillators above express the radial shear in terms of circulation. To determine which further exact families such a relation permits, we consider constant circulation first and then a radial shear that is a ratio of polynomials in the circulation. Constant circulation gives no further family. If $\omega_{\mathrm c}\ne0$, Eq.~(\ref{five}) forces $S_z=\mathcal T_x=0$ and constant $S_x$; the remaining balance is the steady-circulation reversal. If $\omega_{\mathrm c}=0$, the equations reduce to the growing shear--toroidal-field orbit and its stationary limits. When $b_{\mathrm p}=0$, finite circulation gives the toroidal-rotation family.

For nonconstant circulation, suppose $S_x$ depends only on $\omega_{\mathrm c}$ and is a ratio of two polynomials with any common factors cancelled. A zero of the remaining denominator produces a leading term that cannot be cancelled by any other term in Eq.~(\ref{five}), so the denominator must be constant. Powers above $\omega_{\mathrm c}^2$ also produce an uncancelled highest power. The remaining choice has constant dimensionless coefficients $c_0,c_1,c_2$:
\begin{equation}
\frac{S_x}{\Omega}
=c_2\left(\frac{\omega_{\mathrm c}}{\Omega}\right)^2
+c_1\frac{\omega_{\mathrm c}}{\Omega}+c_0.
\end{equation}
Substitution in Eq.~(\ref{five}) and matching equal powers of $\omega_{\mathrm c}$ gives
\begin{equation}
(3c_2-1)(2c_2-1)=0,\qquad
c_1(1-2c_2)=0.
\end{equation}
The choice $c_2=1/3$ gives $c_1=0$ and $c_0=3b_{\mathrm p}^2/\Omega^2-2$, which is the oscillating-circulation cycle. The other choice, $c_2=1/2$, requires $c_1^2=-b_{\mathrm p}^2/\Omega^2$ and therefore gives no real magnetized solution. When $b_{\mathrm p}=0$, it becomes the hydrodynamic circulation--shear oscillator. Thus Eq.~(\ref{elliptic}) is the only real magnetized cycle in which the shear is a ratio of polynomials in the circulation.

Cases in which selected variables remain zero give only the growing magnetic and hydrodynamic families above. Except possibly at isolated values of $b_{\mathrm p}/\Omega$, every conserved polynomial of degree at most four in the five shear--tension variables is a linear combination of $1,C_1,C_2,C_1^2,C_1C_2,C_2^2$. Exact families in which several variables evolve independently may still exist.

\section{Longer period reversal near strong shear}\label{app:long_reversal}
The 11-year dynamo wave need not be the only magnetic cycle operating in the NSSL. A region with stronger shear could support an additional, more slowly reversing wave. We examine its field and flow requirements, the robust relation between an 88-year reversal and a near-203-year modulation, and whether its fast disturbance could reproduce the observed 11-year cycle.

\subsection{Field and flow constraints in the NSSL}
For $\Omega/(2\pi)=443$ nHz, an 88-year reversal gives $\omega_{\mathrm c}=5.7\times10^{-10}\,\mathrm{s^{-1}}$ and $B_{\phi,\max}/B_{\mathrm{lat},\max}\simeq4921q$. At fixed shear, density, and speed, the poloidal amplitude is unchanged from the 11-year wave, but the toroidal amplitude is eight times larger. Kilogauss toroidal fields at shallow densities and speeds of $8$--$12\,\mathrm{m\,s^{-1}}$ therefore favor $q$ near 2, where both amplitudes decrease as $\sqrt{2-q}$. At 2.8 Mm, Model S gives $\rho=2.6\times10^{-5}\,\mathrm{g\,cm^{-3}}$; a stable example is
\begin{equation}
\begin{aligned}
\frac{T_{\mathrm B}}2&=88\,\mathrm{yr},\quad q=1.997,\quad V=8\,\mathrm{m\,s^{-1}},\\
\frac{B_{\phi,\max}}{B_{\mathrm{lat},\max}}&\simeq9800,\\
B_{r,\max}=B_{\mathrm{lat},\max}&\simeq0.56\,\mathrm G,\quad
B_{\phi,\max}\simeq5.5\,\mathrm{kG}.
\end{aligned}
\label{solar_long_reversal}
\end{equation}
At this density, speeds of $6$--$12\,\mathrm{m\,s^{-1}}$ and toroidal amplitudes of $5$--$9$ kG require correlated ranges $q=1.986$--$1.999$ and $0.51$--$0.92$ G poloidally. The corresponding $q_r\simeq2.3$ at $20^\circ$ overlaps enhanced-shear estimates, but simultaneous local field, flow, and shear measurements are lacking. The representative toroidal field is four times the conditional seismic peak of $1.4\pm0.2$ kG near 2.8 Mm~\cite{Baldner}; near $q=2$, a 1-G poloidal field necessarily accompanies about 9.8 kG toroidally. The signed magnetic period and full fluid circuit would be 176 and 352 yr, neither directly measured.

\subsection{Slow and fast disturbances}
The example above has $T_-\simeq203$ yr and $T_+\simeq1.3$ yr. Across $q=1.8$--$1.999$, the slow period changes only from approximately 207 to 203 yr, or $203.2$--$203.5$ yr within the field-compatible range. For $(\omega_{\mathrm c}/\Omega)^2\ll2-q\ll1$, Eq.~(\ref{solar_modulation_periods}) gives
\begin{equation}
T_-\simeq\frac{2T_{\mathrm B}}{\sqrt3}
\simeq203\,\mathrm{yr},
\qquad T_{\mathrm B}=176\,\mathrm{yr}.
\label{app:long_wave_asymptote}
\end{equation}
Once the 88-year reversal is specified, this asymptotic relation gives the near-203-year modulation without precise tuning of the shear. Exactly 208 yr instead requires $q\simeq1.7604$; at 1 Mm, a speed of $5.2\,\mathrm{m\,s^{-1}}$ then gives poloidal and toroidal amplitudes of 1 G and $8.7$ kG.

An 11-year fast disturbance is more restrictive. Substituting $\omega_{\mathrm c}=\pi/T_{\mathrm B}$, $b_{\mathrm p}^2=(2-q)\omega_{\mathrm c}^2/q$, and $\omega_+=2\pi/T_+$ into Eq.~(\ref{fixed_magnetized_dispersion}), with $\omega_-^2=A_0/\omega_+^2$, gives
\begin{equation}
\begin{aligned}
2-q&\simeq4.23\times10^{-5},\\
T_+&=11\,\mathrm{yr},\qquad T_-\simeq203.1\,\mathrm{yr},\\
\frac{B_{\phi,\max}}{B_{\mathrm{lat},\max}}&\simeq9840.
\end{aligned}
\label{solar_three_frequencies}
\end{equation}
At the same 2.8-Mm density, speeds of $8$--$12\,\mathrm{m\,s^{-1}}$ give $0.066$--$0.099$ G poloidally and $0.65$--$0.98$ kG toroidally. Since $T_+\simeq T_{\mathrm{orb}}/\sqrt{2-q}$, fast periods of $9$--$13$ yr require $2-q\simeq3.0\times10^{-5}$--$6.3\times10^{-5}$, far narrower than observational shear resolution. Therefore, reconciling the $11$ yr cycle with the fast mode of the $88$ yr reversal orbit requires fine-tuning the shear. 

Increasing the fast-mode amplitude does not recover the observed solar cycle within realistic flow limits. Even allowing speeds of approximately $0$--$20\,\mathrm{m\,s^{-1}}$ around a $10\,\mathrm{m\,s^{-1}}$ reference flow gives only a $0.1\%$ toroidal-envelope semi-amplitude and a few-percent rapid field variation, while reversals remain near 88 yr. Nonlinear integrations with the steady wave's $C_1,C_2$ held fixed confirm this behavior. More decisively, $\dot\phi_{\mathrm B}=2\omega_{\mathrm c}$ requires at least 44 yr for a complete poloidal reversal under this speed limit; reversal in 11 yr would require an interval-averaged speed of $80\,\mathrm{m\,s^{-1}}$. The fast disturbance therefore cannot supply the observed 11-year magnetic reversal.

The failure of the fast disturbance to reproduce the 11-year polarity cycle does not exclude coexistence of separate 11- and 88-year dynamo waves in regions with different shear. Their sum is not an exact non-linear solution, but limited spatial overlap could weaken their interaction. At a fixed position, the longer wave's signed field repeats after 176 yr, whereas its toroidal-field magnitude and local magnetic energy repeat after 88 yr, providing a possible contribution to activity variations. Century-scale changes in hemispheric asymmetry offer some motivation: Pulkkinen et al.~\cite{Pulkkinen1999} proposed an oscillating magnetic component to explain an approximately 90-year displacement of the activity belts, while Mursula~\cite{Mursula2023} proposed a persistent magnetic field whose position oscillates over approximately 210 yr. Neither establishes an 88-year polarity reversal. 

\subsection{The corresponding tachocline case}
At tachocline values $\Omega/(2\pi)=410$ nHz, $\rho=0.19\,\mathrm{g\,cm^{-3}}$, and equatorward speed $4\,\mathrm{m\,s^{-1}}$, an 88-year reversal at $q=0.31$ requires about 2 MG toroidally and is unstable, with a 36-year growth time. A stable wave with 11- and 203-year disturbances instead requires $2-q\simeq4.94\times10^{-5}$; speeds of $3$--$4\,\mathrm{m\,s^{-1}}$ give $2.3$--$3.1$ G poloidally and $21$--$28$ kG toroidally. These toroidal fields lie below the geometry-dependent seismic upper bound but above the $1.4$--$6.5$-kG reconstruction discussed in the main text. Exactly 208 yr requires $q\simeq1.7604$ and much stronger toroidal fields, $1.4$--$1.8$ MG. Recent tachocline-width measurements do not establish the required cylindrical shear~\cite{BasuTach2026,KorzennikEffDarwich2026}, which must coexist with the adopted density and flow.

\bibliography{references}
\end{document}